\documentclass[journal]{IEEEtran}

\usepackage{cite}
\usepackage{amsmath,amssymb,amsfonts}
\usepackage{algorithmic}
\usepackage{textcomp}
\usepackage{xcolor}
\usepackage{booktabs}
\usepackage{array}
\usepackage{siunitx}
\usepackage{subcaption}
\usepackage{graphicx}
\usepackage{dblfloatfix}
\usepackage{placeins}
\usepackage{cuted}
\usepackage[colorlinks=true,citecolor=blue,linkcolor=blue,urlcolor=blue]{hyperref}
\usepackage{orcidlink}
\usepackage[utf8]{inputenc}
\usepackage[T1]{fontenc}

\title{Local Differential Privacy for Molecular Communication Networks}

\author{Melih~\c{S}ahin \orcidlink{0009-0000-6813-3297}, ~\IEEEmembership{Graduate Student Member,~IEEE,}
Ozgur~B.~Akan \orcidlink{0000-0003-2523-3858},~\IEEEmembership{Fellow,~IEEE}
\thanks{Melih \c{S}ahin is with the Centre for neXt Communications (CXC), Department of Engineering, University of Cambridge, CB3 0FA Cambridge, U.K. (e-mail: ms3195@cam.ac.uk).}
\thanks{Ozgur B. Akan is with the Centre for neXt Communications (CXC), Department of Engineering, University of Cambridge, CB3 0FA Cambridge, U.K., and also with the Center for neXt Communications (CXC), Department of Electrical and Electronics Engineering, Ko\c{c} University, 34450 Istanbul, T\"urkiye (e-mail: oba21@cam.ac.uk).}
\thanks{This work was supported in part by the AXA Research Fund (AXA Chair for Internet of Everything at Koç University).}
}

\begin{document} \bstctlcite{IEEEexample:BSTcontrol} \maketitle \begin{abstract}
Molecular communication (MC) enables information exchange in nanoscale sensor networks operating in biological environments, yet privacy remains largely unaddressed. We integrate local differential privacy (LDP) into diffusion-based MC by privatizing each user’s measurement at the transmitter and conveying the resulting randomized report over the MC channel. To our knowledge, this is the first systematic LDP implementation for diffusion-based MC, enabling privacy-preserving aggregate data analysis for in-body health monitoring and other population-scale sensing applications. We benchmark major LDP mechanisms under a realistic channel model. Simulation results show that Optimized Local Hashing (OLH) achieves the lowest average $\ell_1$ distribution-estimation error under the MC channel when channel resources are sufficient and the number of possible user values (alphabet size) $k$ is moderate to large, whereas k-ary Randomized Response (KRR) and, for privacy budget $\epsilon=1$, Binary Local Hashing (BLH), which have shorter reports, are more robust as the MC transmission quality deteriorates. We further propose RLIM-LDP, which combines run-length-limited ISI-mitigation (RLIM) coding with LDP coding. Extensive simulation results demonstrate that RLIM-LDP improves end-to-end reliability and reduces the final distribution-estimation error when time and molecule resources are limited.
\end{abstract} 

\section{Introduction}

Diffusion-based molecular communication (MC) uses the controlled release, propagation, and detection of molecules to convey information in fluidic media where electromagnetic communication can be unreliable or impractical \cite{ozgur_hoca_survey}. This makes MC a natural candidate for communication among nanoscale devices in diffusive biological environments, such as the human vascular system.

Privately transmitting individual information at population scale is an important task, especially when the data contains highly personal information. In MC, security and privacy have been highlighted as major open challenges, with early work outlining MC-specific attack methods and noting that conventional cryptographic defenses may be impractical at the nano/micro-scale \cite{mc_privacy_1}. Physical-layer security has also been studied for MC: the secrecy capacity and secure distance of a diffusion-based link with an eavesdropper are derived in \cite{mucchi_secrecy}, physical-layer counterattack strategies for the Internet of Bio-Nano Things (IoBNT) \cite{iobnt} are proposed in \cite{huang_counterattack}, and \cite{qiu} reviews physical-layer security in the molecular Internet of Nano-Things. Recent efforts also explore information-theoretic secure communication viewpoints for Poisson models motivated by medical MC \cite{mc_privacy_2}. These works protect the transmitted message from an external adversary while the receiver and the data collector are trusted. However, beyond defending against external adversaries, many in-body health-monitoring scenarios require a stronger notion of privacy in which even the data collector should not be able to reliably infer any individual’s sensitive value; this notion is complementary to physical-layer security and, to our knowledge, has not been studied for MC before.

A rigorously-defined way to privately send information, in the sense that privacy is protected even from the receiver itself, is through local differential privacy (LDP) \cite{2-oue-4-olh-7-blh}. The output of each user is randomized based on the LDP mechanism used. This gives the user plausible deniability and thus ensures privacy. Once sufficiently many users send privatized reports, the receiver can estimate the underlying distribution. A central purpose of LDP is therefore to enable aggregate data analysis while preserving individual privacy. In this work, we focus on the fundamental LDP problem of estimating a $k$-ary probability distribution when $N$ users report privatized messages to a central server. For instance, each of these $N$ subjects may be a human, and the report may be the presence or absence of a disease, where the aim is to estimate the prevalence of that disease among the general public, without endangering personal privacy.

Performance comparisons of major LDP schemes are well documented in the literature \cite{2-oue-4-olh-7-blh,p2046-cormode}. However, in the resource-constrained setting of MC, the communication cost (both the length of the transmitted binary string and the number of $1$-bits it contains) can significantly affect communication quality \cite{MC_channel_coding}. Since MC has strong potential for enabling in-vivo body communication, it is therefore important to compare existing LDP schemes specifically in the context of an MC channel. To the best of our knowledge, this is the first work to provide such a comparison.

The remainder of this paper is organized as follows. Section~\ref{sec:channel} introduces the diffusion-based MC channel model. Section~\ref{sec:ldp} details the LDP architecture and the LDP mechanisms evaluated in this study. Section~\ref{sec:sim} presents our simulation framework and compares these privacy schemes under realistic MC channel conditions. Section~\ref{sec:mc_channel_coding_ldp} proposes the RLIM-LDP framework, which combines run-length-limited inter-symbol interference (ISI)-mitigation (RLIM) coding \cite{MC_channel_coding} with LDP coding. Section~\ref{sec:conclusion} concludes the paper.

\section{Molecular Communication Channel}\label{sec:channel}

\begin{figure}[t]
\centering
\includegraphics[width=0.4\textwidth]{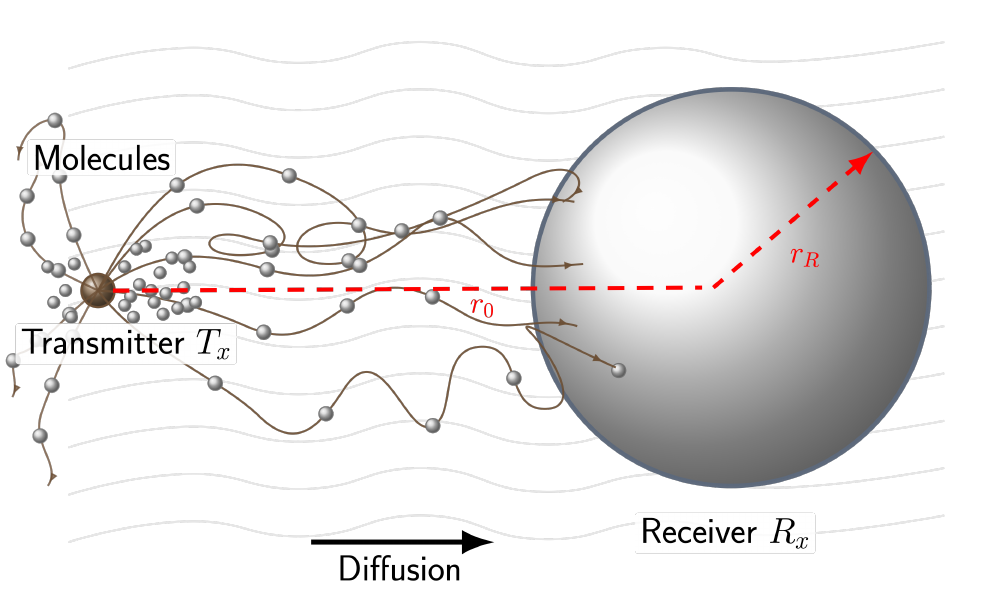}
\caption{MC Channel Model}
\label{fig:channel}
\end{figure}

The classical SISO (single-input single-output) molecular communication system is shown in Fig.~\ref{fig:channel}. The information is transmitted using binary concentration shift keying (BCSK) modulation \cite{BCSK} with a fully absorbing spherical receiver: The time axis is divided into equally long signal intervals (i.e., time slots). In each signal interval, a single bit (either a 0-bit or a 1-bit) is transmitted. If a 1-bit is to be transmitted, at the start of the corresponding signal interval, $M$ molecules are released from the transmitter and diffuse via Brownian motion \cite{MC_main}. If a 0-bit is to be transmitted, no molecules are emitted from the transmitter. Some of these $M$ molecules may be detected (i.e., absorbed) by the receiver within the current interval, within ensuing intervals, or not be detected at all. The receiver counts the number of molecules it absorbs within each signal interval. Therefore, in a SISO MC model, for a binary sequence of length $n$, we obtain a nonnegative integer sequence of length $n$.

The probability of a single molecule released from the transmitter getting absorbed by the receiver within time $t$ (in $\mathrm{s}$) is given by \cite{channel_characteristics}: \begin{equation}\label{eq:F}
F(t) = \frac{r_R}{r_0}\,\mathrm{erfc}\!\left( \frac{r_0 - r_R}{\sqrt{4Dt}} \right),
\end{equation} where $D$ is the diffusion coefficient (in $\mu\mathrm{m}^2/\mathrm{s}$), $r_R$ is the receiver radius (in $\mu\mathrm{m}$), and $r_0$ denotes the transmitter--receiver distance (center-to-center) in $\mu\mathrm{m}$.

\begin{figure*}[!t]
\centering
\includegraphics[width=\textwidth]{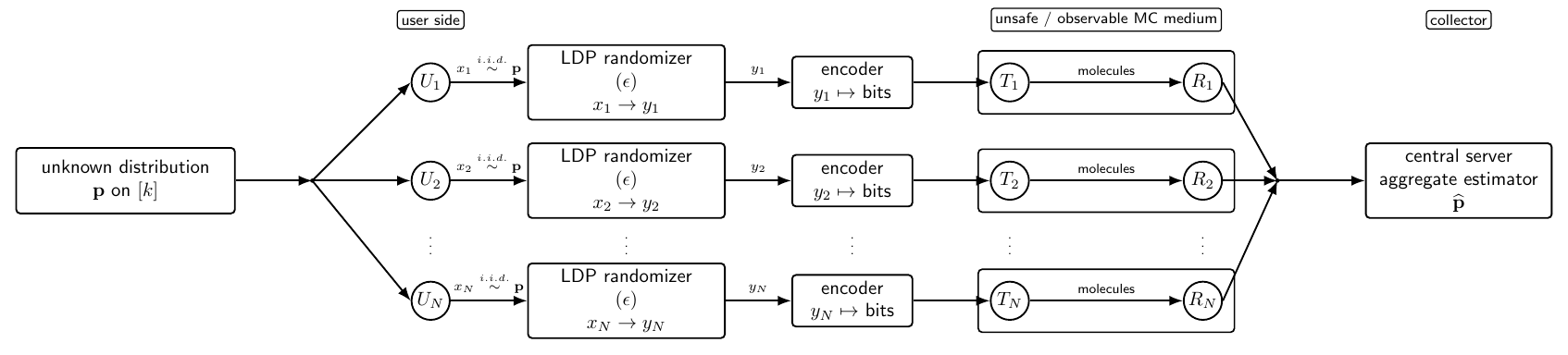}
\caption{Network architecture for local differential privacy in molecular communication}
\label{fig:architecture}
\end{figure*}

The $i$-th channel coefficient is defined as the probability that a single molecule released at the start of the $j$-th signal interval gets absorbed within the $(j+i-1)$-th signal interval, and it is given by $p_i = F(i\cdot t_s) - F((i-1)\cdot t_s)$ for signal interval duration $t_s$ (in $\mathrm{s}$) \cite{ozgur_hoca_survey}.

For a transmitted 1-bit, each of the $M$ emitted molecules falls into one of the following mutually exclusive outcomes: it is absorbed in the 1st, 2nd, \dots, $I$-th subsequent signal interval, or it is not absorbed within the modeled memory window (i.e., it arrives later than $I$ intervals, or it never arrives). Thus, for a single emission event, the vector of molecule counts across these outcomes follows a multinomial distribution \cite{multinomial}: \begin{equation}
\big(X_1,\ldots,X_I,X_{I+1}\big)\sim \mathrm{Multinomial}\!\left(M;\, p_1,\ldots,p_I,\, p_{I+1}\right),
\end{equation} where $X_i$ denotes the number of molecules absorbed in the $i$-th interval after emission, and $p_{I+1}=1-\sum_{i=1}^{I}p_i$ collects the probability mass of molecules that are not absorbed within the first $I$ intervals. The constant $I$ is introduced to simulate the MC channel under constrained computational resources by truncating the channel memory. When $I$ is chosen sufficiently large, this truncation causes only a slight deviation from the most accurate (i.e., $I=n$) model, since the coefficients $p_i$ rapidly decay as $i$ increases, and the remaining tail probability is small (it is quantified in Section~\ref{sec:sim_mech}). Then, for a transmitted binary sequence $\{b_j\}_{j=1}^{n}$ with $b_j\in\{0,1\}$, the received molecule count in the $h$-th interval can be modeled as \begin{equation}\label{eq:Nh}
N_h=\sum_{i=1}^{\min\{h,I\}} b_{h-i+1}\,X_i^{(h-i+1)} + W,
\end{equation} where $X_i^{(m)}$ is the multinomial component \cite{multinomial} corresponding to molecules emitted in interval $m$ and absorbed $i$ intervals later, and $W$ denotes receiver counting noise. In our simulations, additive Gaussian noise with variance $\sigma^2$ and mean $0$, rounded to the nearest integer, has been used. \section{LDP System Model}\label{sec:ldp} In a typical LDP model \cite{2-oue-4-olh-7-blh}, it is assumed that there are $N$ entities sending information to a central server for aggregate data analysis. While in the classical definition of differential privacy this central server is assumed to be trusted and secure, LDP assumes that even the central server may not be secure (i.e., it may be curious or untrusted). Aggregate data processing is used for many different purposes, such as machine learning data collection, large-scale data analytics, public health monitoring, and privacy-preserving surveys or censuses, where only population-level frequencies are needed.

A central benchmark for evaluating an LDP mechanism is the problem of $k$-ary probability distribution estimation \cite{2-oue-4-olh-7-blh}, in which each user holds a categorical value $X \in [k] = \{1,2,\ldots,k\}$ drawn i.i.d.\ from an unknown distribution $\boldsymbol{p} \in S_k = \{ \boldsymbol{p}\in\mathbb{R}^k:\ p_i \ge 0,\ \sum_{i=1}^k p_i = 1 \}$, where $S_k$ denotes the probability simplex. Throughout, whenever a binary or base-$g$ representation of a symbol $x\in[k]$ is needed, the integer $x-1\in\{0,\ldots,k-1\}$ is encoded, so that $\lceil\log_2 k\rceil$ bits (or $\lceil\log_g k\rceil$ base-$g$ digits) suffice even when $k$ is a power of the base. An LDP mechanism is defined by an output alphabet $Y$ and a conditional distribution (equivalently, a probabilistic matrix) $Q \in [0,1]^{k\times |Y|}$, $Q_{x,y} = \Pr(Y=y\,|\,X=x)$: for each input $x\in[k]$, the mechanism outputs a randomized report $y\in Y$ according to the $x$-th row of $Q$.

An LDP mechanism is $\epsilon$-locally differentially private iff for all $x,x'\in[k]$ and for all $y\in Y$, \begin{equation}
\frac{\Pr(Y=y\,|\,X=x)}{\Pr(Y=y\,|\,X=x')} \le e^{\epsilon},
\end{equation} where $\epsilon\ge0$ \cite{2-oue-4-olh-7-blh}. Smaller values of $\epsilon$ provide a higher degree of privacy; $\epsilon=0$ corresponds to perfect privacy, but then the privatized reports carry no useful information, so meaningful estimation of $\boldsymbol{p}$ is not possible.

We demonstrate our privacy model for MC in Fig.~\ref{fig:architecture}. In this architecture, there are $N$ subjects (e.g., human users), each equipped with an in-body sensor--gateway pair. For subject $i\in[N]$, the in-body sensor node $T_i$ (equipped with sensing and molecular transmission capability) observes a value $x\in[k]$, privatizes it by randomly choosing $y\in Y$ according to $\Pr(Y=y\,|\,X=x)$, and transmits a binary representation of $y$ over a diffusion-based MC link. The receiver $R_i$ detects (absorbs/counts) the molecular signal and then forwards the decoded report to a distant central server using classical electromagnetic (EM) communication. We assume that this second channel is error-free, since it is a conventional digital link with error-control coding and retransmissions whose residual error rate is orders of magnitude below the bit error rates (BERs) of the MC link; all transmission errors in our model thus originate from the MC link $T_i\!\to\!R_i$, and, as shown in Section~\ref{sec:sim_mech}, the privacy guarantee does not depend on this assumption.

While EM links are feasible for many implants, direct EM transmission from $T_i$ may be unreliable or undesirable in the regime we consider due to strong tissue-dependent attenuation, antenna size constraints at the nano/micro-scale (wavelength-scale effects), and strict power/thermal budgets \cite{mi14071472}. In such short-range, fluidic biological environments (e.g., intravascular or interstitial media), chemically mediated diffusion can be a more compatible in-body physical layer; our focus is therefore the regime where an MC channel $T_i\!\to\!R_i$ is necessary.

It may be argued that privatization could instead be performed at the gateway side $R_i$. However, we do not assume the MC channel is trusted: adversarial nano/micro-scale entities in the environment may observe the molecular signal, and the gateway $R_i$ itself may be untrusted in some deployments. Therefore, to provide an end-to-end local privacy guarantee independent of the receiver, privatization is performed at the sensor-transmitter side $T_i$. We now present the standard LDP schemes used in the literature as follows:

\subsection{The $k$-ary Randomized Response (KRR)}\label{sec:krr} In KRR \cite{KRR}, the output alphabet is the same as the input, i.e., $Y=[k]$. Given a symbol $x\in[k]$, the user reports $Y=x$ with probability $p=\frac{e^{\epsilon}}{e^{\epsilon}+k-1}$, and otherwise a $y\neq x$ chosen uniformly over $[k]\setminus\{x\}$. The server receives $N$ reports $Y_1,\dots,Y_N$ and forms the empirical histogram $\hat{f}_j=\frac{1}{N}\sum_{i=1}^N \mathbf{1}\{Y_i=j\}$. An unbiased estimate (i.e., $\mathbb{E}[\hat{p}_j]=p_j$ for all $j\in[k]$) of the underlying distribution is then $\hat{p}_j=\frac{\hat{f}_j-q}{p-q}$ with $q=\frac{1}{e^{\epsilon}+k-1}$.

\subsection{Basic RAPPOR (Symmetric Unary Encoding)}\label{sec:rappor} Basic RAPPOR \cite{basic_rappor} represents the input $x\in[k]$ as a unary (i.e., containing only one 1-bit) vector $\mathbf{v}\in\{0,1\}^k$ with $v_x=1$ and $v_j=0$ for $j\neq x$. The user then flips each bit independently: a $1$ stays $1$ with probability $p$ (and flips to $0$ with probability $1-p$), while a $0$ flips to $1$ with probability $q$. In basic RAPPOR, $p=\frac{e^{\epsilon/2}}{e^{\epsilon/2}+1}$ and $q=\frac{1}{e^{\epsilon/2}+1}$. Let $\mathbf{z}_i\in\{0,1\}^k$ be the privatized vector from user $i$, and define the average $\bar{z}_j=\frac{1}{N}\sum_{i=1}^N z_{i,j}$. The unbiased estimator is $\hat{p}_j=\frac{\bar{z}_j-q}{p-q}$. \subsection{Optimized Unary Encoding (OUE)}\label{sec:oue} OUE \cite{2-oue-4-olh-7-blh} uses the same unary representation and estimator as basic RAPPOR, but chooses $p=\frac{1}{2}$ and $q=\frac{1}{e^{\epsilon}+1}$ to reduce the estimation variance.

\subsection{Binary Local Hashing (BLH)}\label{sec:blh} In BLH \cite{2-oue-4-olh-7-blh}, each $x\in[k]$ is identified with the binary representation of $x-1$, $\mathbf{x}\in\{0,1\}^d$, where $d=\lceil \log_2 k \rceil$. The user draws a random hash vector $\mathbf{r}\in\{0,1\}^d$ uniformly, computes the hashed bit $t = \langle \mathbf{r},\mathbf{x}\rangle \bmod 2 = \big( \sum_{i=1}^d r_i x_i \big) \bmod 2$, and applies binary randomized response to $t$: the reported bit equals $t$ with probability $p=\frac{e^{\epsilon}}{e^{\epsilon}+1}$, and is flipped otherwise. The transmitted report is the pair $(\mathbf{r},Y)$, where $Y\in\{0,1\}$ is the randomized reported bit.

At the server, for each candidate $v\in[k]$ with binary form $\mathbf{v}$, let $c_v= \frac{1}{N} \sum_{i=1}^{N}\mathbf{1}\{Y_i=\langle \mathbf{r}_i,\mathbf{v}\rangle \bmod 2\}$ be the fraction of reports matching the predicted hash. Then $\mathbb{E}[c_v]=p_v\,p+(1-p_v)\cdot \tfrac{1}{2}$, where $p_v=\Pr(X=v)$: for $X\neq v$ we have $\mathbf{x}\oplus\mathbf{v}\neq \mathbf{0}$, so with uniform $\mathbf{r}$ the hashed bits of $\mathbf{x}$ and $\mathbf{v}$ match with probability $\tfrac{1}{2}$, and randomized response cannot change this symmetry. Thus an unbiased estimator is $\hat{p}_v=\frac{c_v-1/2}{p-1/2}$.

\subsection{Optimized Local Hashing (OLH)}\label{sec:olh} OLH \cite{2-oue-4-olh-7-blh} is the $g$-ary analogue of BLH: we hash to an alphabet of size $g$ and then apply $g$-ary randomized response. The transmitted report is $(h,Y)$, where $h:[k]\to[g]$ is a randomly chosen hash function and $Y\in[g]$ is the randomized hash output. The crucial requirement is the following symmetry: for a given family of hash functions $H$ and for any two distinct symbols $x\neq v$, \begin{equation}\label{eq:olh_symmetry}
\Pr_{h\in H}\!\big(h(x)=h(v)\big)=\frac{1}{g}.
\end{equation} This is the $g$-ary counterpart of the $\tfrac{1}{2}$ symmetry in BLH. The choice of $g$ determines the estimation variance, $V(g;\epsilon) \propto \frac{(e^{\epsilon}-1+g)^2}{(e^{\epsilon}-1)^2\,(g-1)}$, $g\ge 2$, which is minimized at the continuous optimum $g=e^\epsilon+1$. Universal hash families $H$ satisfying \eqref{eq:olh_symmetry} are known to exist whenever $g$ is a prime power, because a finite field with $g$ elements (denoted $\mathrm{GF}(g)$) exists in that case. Such constructions typically use arithmetic over $\mathrm{GF}(g)$ (e.g., polynomial-based maps), which is more involved to implement on a severely constrained nano/micro-scale transmitter. In our MC setting, we therefore use a modulo-based construction that achieves \eqref{eq:olh_symmetry} using only integer arithmetic, which requires $g$ to be \emph{prime} (so that $\mathbb{Z}_g$ is a field): Let $g_0=\lfloor e^{\epsilon}\rfloor+1$, take the nearest primes $g^-=\max\{q\le g_0:\ q\ \text{prime}\}$ and $g^+=\min\{q>g_0:\ q\ \text{prime}\}$, and set $g=\arg\min_{q\in\{g^-,g^+\}} V(q;\epsilon)$.

Then, let $m=\lceil \log_g k\rceil$ and represent each $x\in[k]$ by the length-$m$ base-$g$ vector of $x-1$, $\mathbf{x}\in\{0,1,\ldots,g-1\}^m$. The user samples $\mathbf{r}\in\{0,1,\ldots,g-1\}^m$ uniformly and defines the hash \begin{equation}\label{eq:olh_hash}
h_{\mathbf{r}}(x)=\langle \mathbf{r},\mathbf{x}\rangle \bmod g
=\Big(\sum_{j=1}^{m} r_j x_j\Big)\bmod g,
\end{equation} which satisfies \eqref{eq:olh_symmetry} whenever $g$ is prime, since for any $x\neq v$ the random variable $\langle \mathbf{r},\mathbf{x}-\mathbf{v}\rangle \bmod g$ is uniform over $[g]$. Given $t=h(x)\in[g]$, the user applies $g$-ary randomized response: $\Pr(Y=t)=p=\frac{e^{\epsilon}}{e^{\epsilon}+g-1}$, and otherwise $Y$ is chosen uniformly from $[g]\setminus\{t\}$, i.e., $\Pr(Y=y)=\frac{1-p}{g-1}$ for $y\neq t$. 

This provides us with $\epsilon$-local privacy. Since our diffusion-based MC model uses binary modulation, before transmitting $(h,Y)\in[g]^{m+1}$, we map it to an integer $z\in \{0,\ldots,g^{m+1}-1\}$ via its base-$g$ expansion, and transmit the fixed-length binary representation of $z$ with $\lceil \log_2(g^{m+1})\rceil$ bits. At the server, for each candidate $v\in[k]$, let $c_v=\frac{1}{N}\sum_{i=1}^{N}\mathbf{1}\{Y_i=h_i(v)\}$. Using \eqref{eq:olh_symmetry}, one obtains $\mathbb{E}[c_v]=p_v \cdot p +(1-p_v) \cdot \frac{1}{g}$; hence an unbiased estimator is $\hat{p}_v=\frac{c_v-1/g}{p-1/g}$.

\subsection{Hadamard Response (HR)}\label{sec:hr} HR \cite{hadamard} achieves strong accuracy guarantees while enabling fast server-side frequency estimation. For brevity, we detail the core HR mechanism optimized for the high-privacy regime ($\epsilon = O(1)$) below; in our simulations for $\epsilon > 1$, we implement the generalized block-structured HR construction detailed in Appendix A of \cite{hadamard}, which maintains sample optimality by dynamically scaling the subset size. Let $K=\left\lceil \log_2(k+1)\right\rceil$, $d=2^K$, and identify each $j\in[d]$ with the $K$-bit binary vector of $j-1$. Define the Hadamard matrix $H\in\{\pm1\}^{d\times d}$ by $H_{j,t}=(-1)^{\langle j,t\rangle}$, where $\langle j,t\rangle$ is the (mod-$2$) inner product of the two $K$-bit vectors. Given input $x\in[k]$, define the “positive-support” index set $S_x = \{\, j\in[d] : H_{x+1,j}=+1 \,\}$ (where the row index $x+1$ skips the all-ones first row of $H$). Note that $|S_x|=d/2$ for every $x$, and $|S_x \cap S_y| = |S_x - S_y| = d/4$ for all $x \neq  y \in [k]$. With probability $p=\frac{e^{\epsilon}}{e^{\epsilon}+1}$, the user samples $Y$ uniformly from $S_x$; with probability $1-p$, uniformly from $[d]\setminus S_x$, i.e., $\Pr(Y=j\mid X=x)=\frac{2p}{d}$ for $j\in S_x$ and $\frac{2(1-p)}{d}$ for $j\notin S_x$. This satisfies $\epsilon$-LDP since the maximum likelihood ratio is $\frac{p}{1-p}=e^\epsilon$.

\captionsetup[subfigure]{labelformat=parens} \captionsetup[subfigure]{justification=centering, singlelinecheck=false}

\begin{figure*}[!t]
\centering
\newlength{\PANELW}
\setlength{\PANELW}{0.25\textwidth}
\newlength{\ROWGAP}
\setlength{\ROWGAP}{0.5mm}
\begin{tabular}{@{}c@{}c@{}c@{}c@{}}
\multicolumn{4}{@{}c@{}}{
  \includegraphics[width=3.8\PANELW]{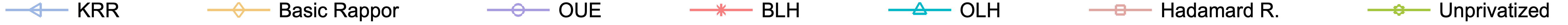}
}\\[-1mm]
  \subcaptionbox{$M=1000$, $t_s=1$ $\mathrm{s}$, $k=16$}{\includegraphics[width=\PANELW]{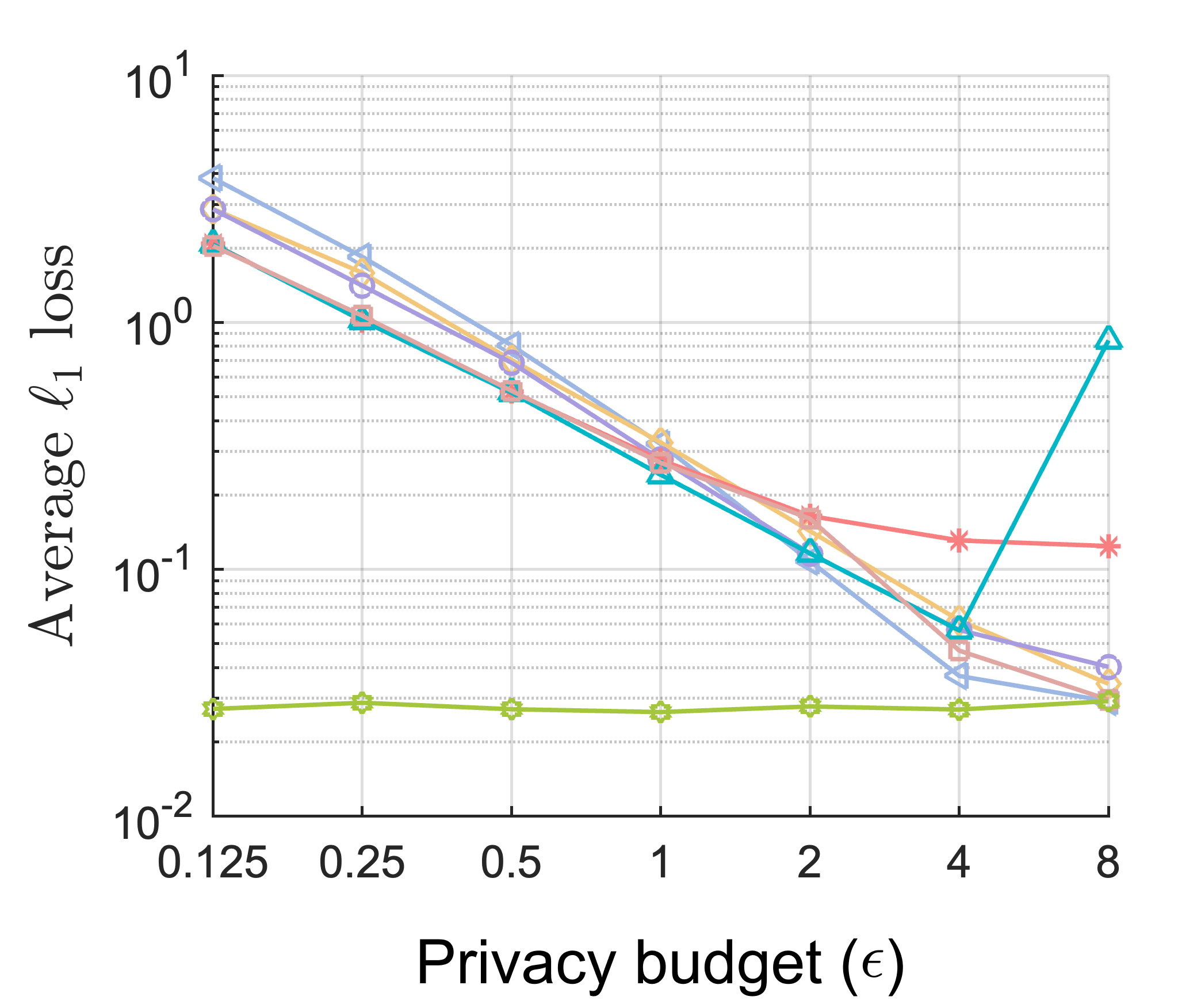}} &
  \subcaptionbox{$\epsilon=1$, $M=1000$, $k=16$}{\includegraphics[width=\PANELW]{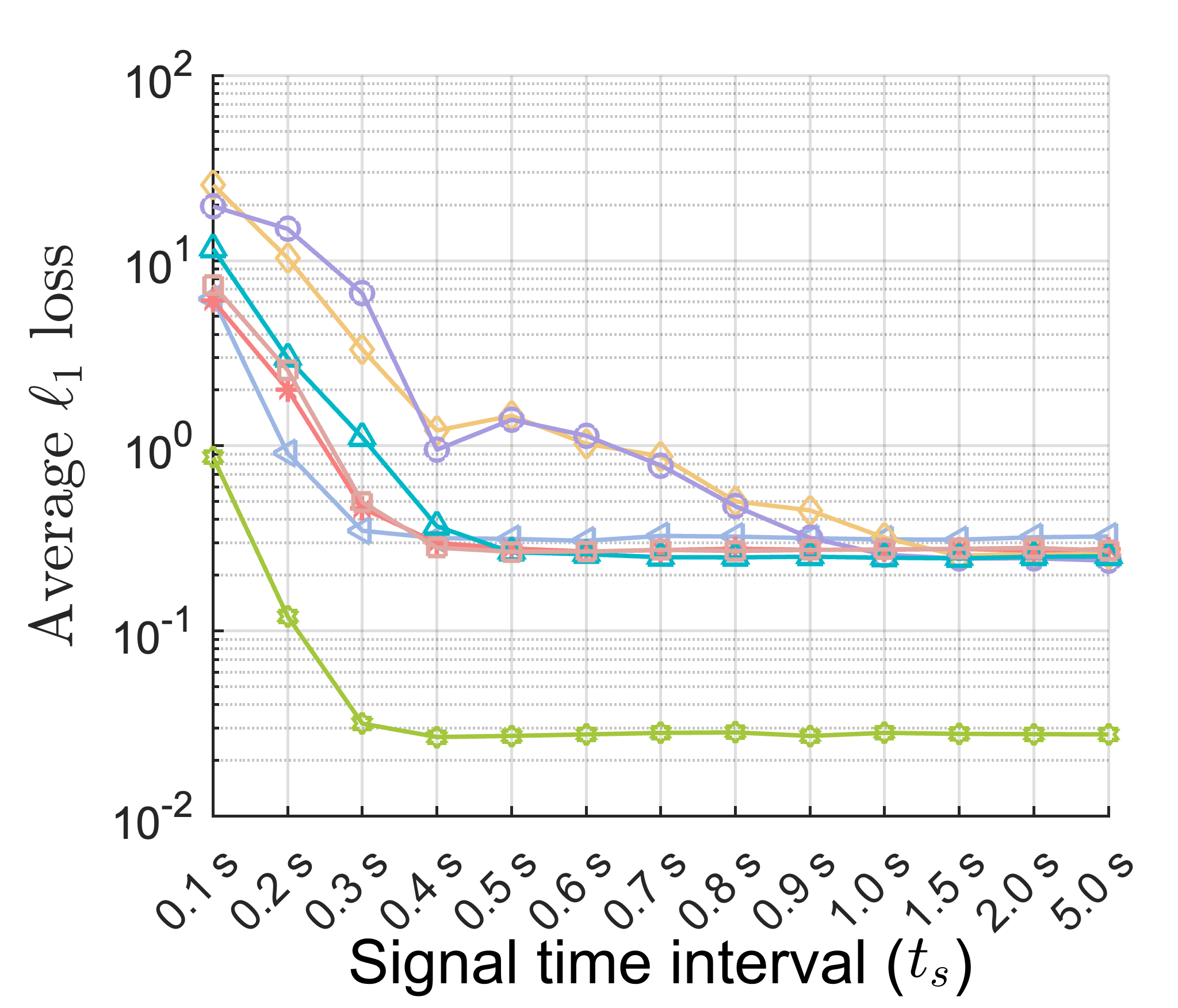}} &
  \subcaptionbox{$\epsilon=1$, $t_s=1$ $\mathrm{s}$, $k=16$}{\includegraphics[width=\PANELW]{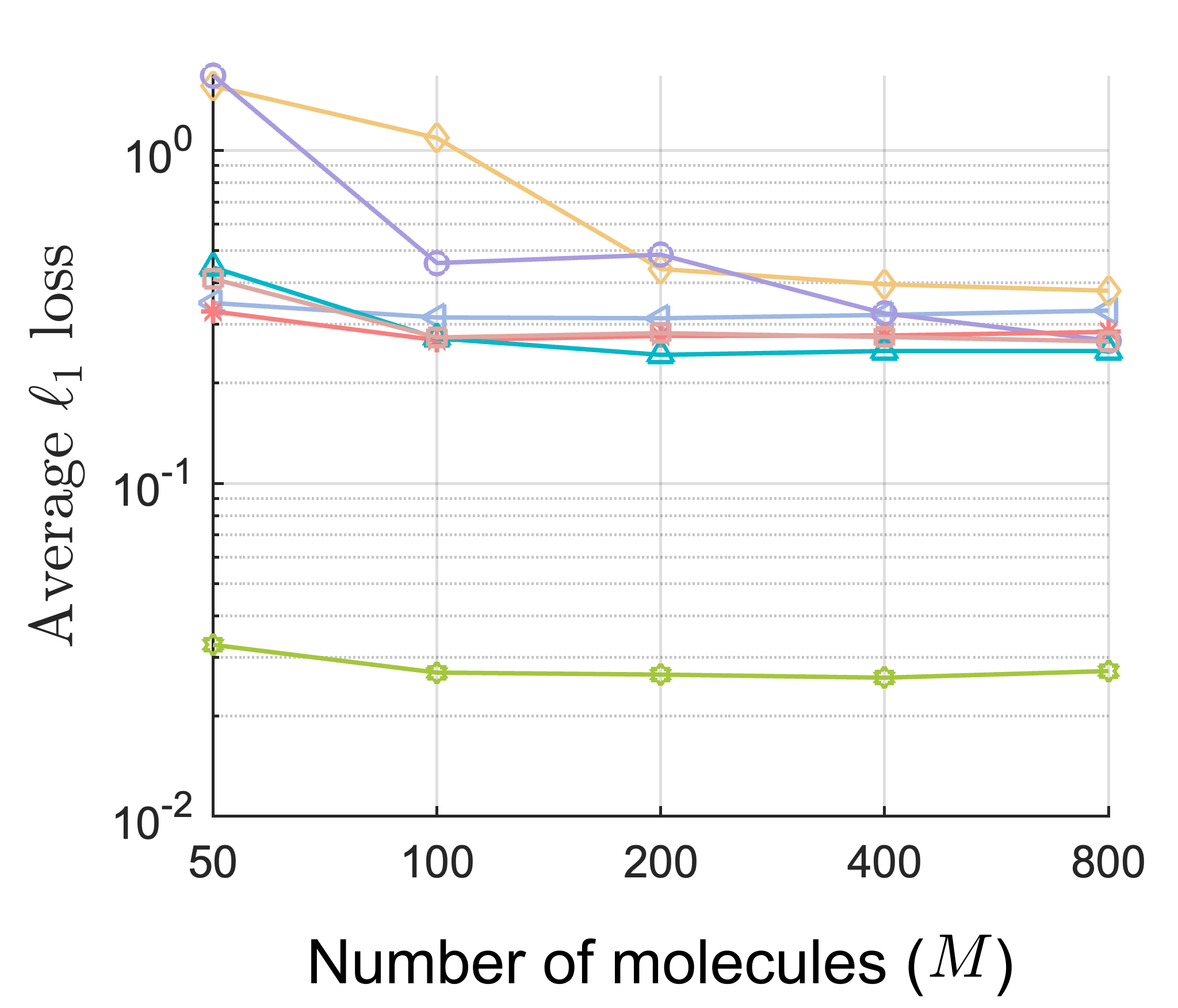}} &
  \subcaptionbox{$\epsilon=1$, $M=1000, t_s=1$ $\mathrm{s}$, $k=16$}{\includegraphics[width=\PANELW]{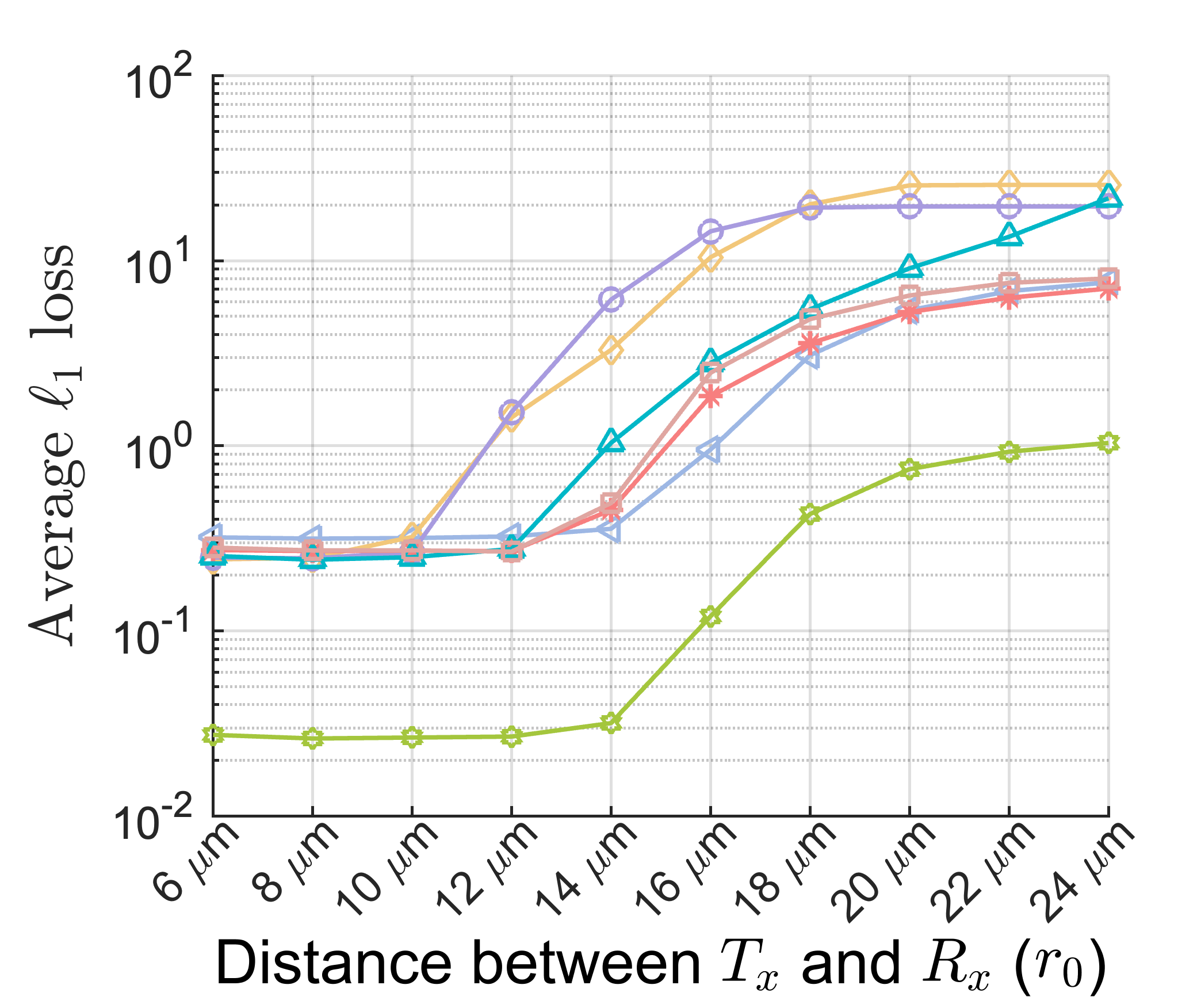}}
  \\[\ROWGAP]
  \subcaptionbox{$\epsilon=1$, $M=1000, t_s=1$ $\mathrm{s}$, $k=16$}{\includegraphics[width=\PANELW]{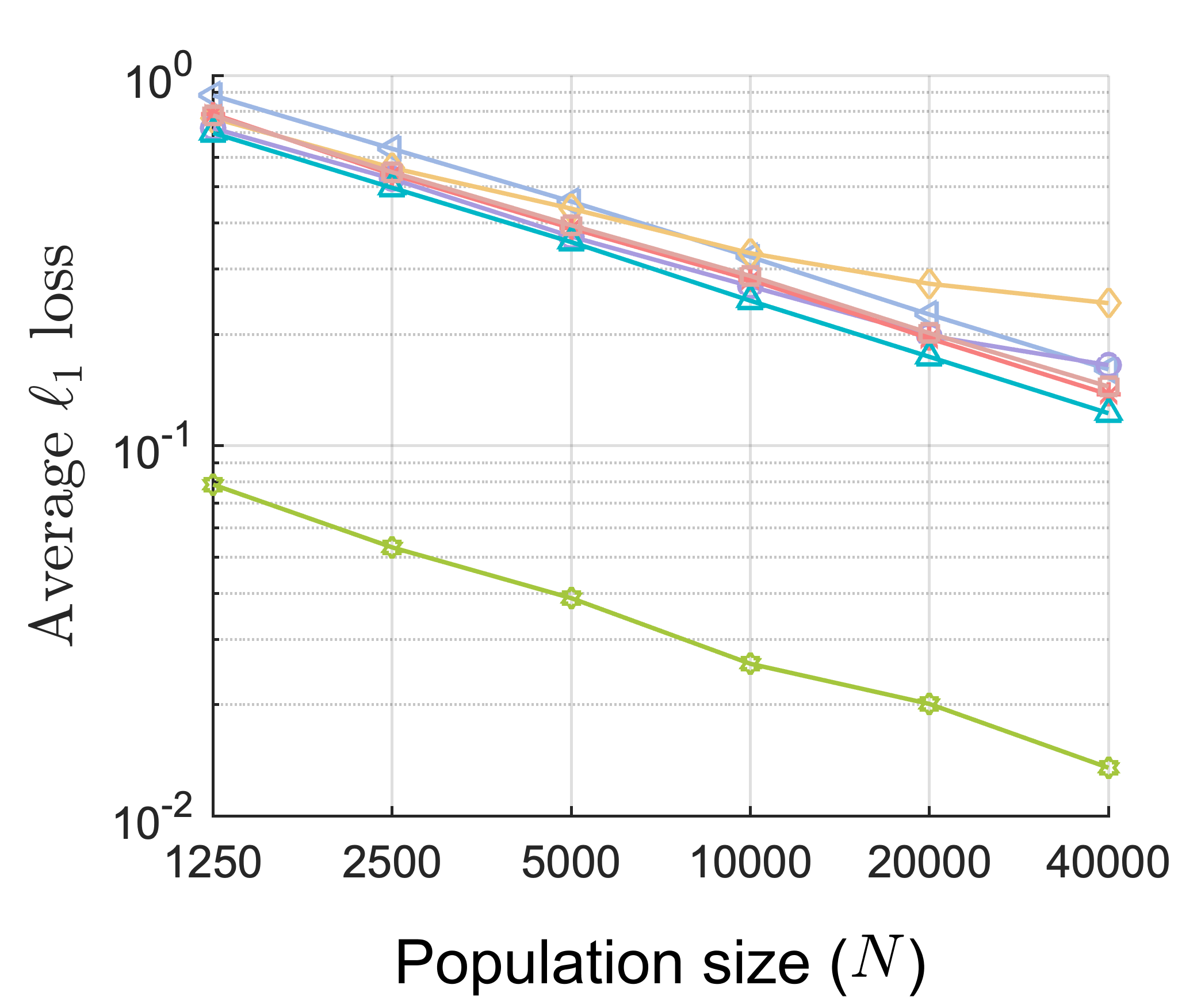}} &
  \subcaptionbox{$\epsilon=1$, $M=100, t_s=1$ $\mathrm{s}$, $k=16$}{\includegraphics[width=\PANELW]{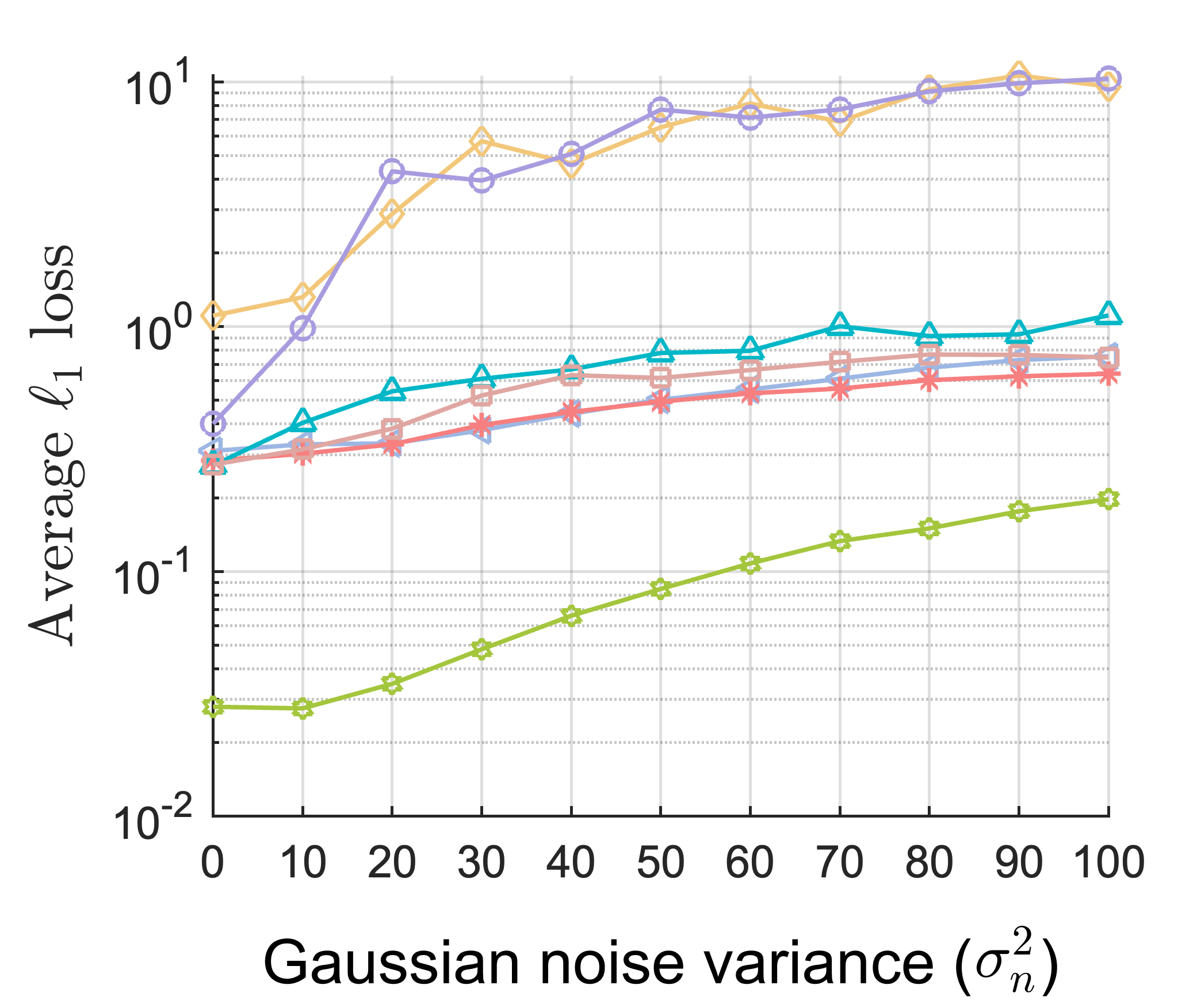}} &
  \subcaptionbox{$\epsilon=1$, $M=1000, t_s=1$ $\mathrm{s}$}{\includegraphics[width=\PANELW]{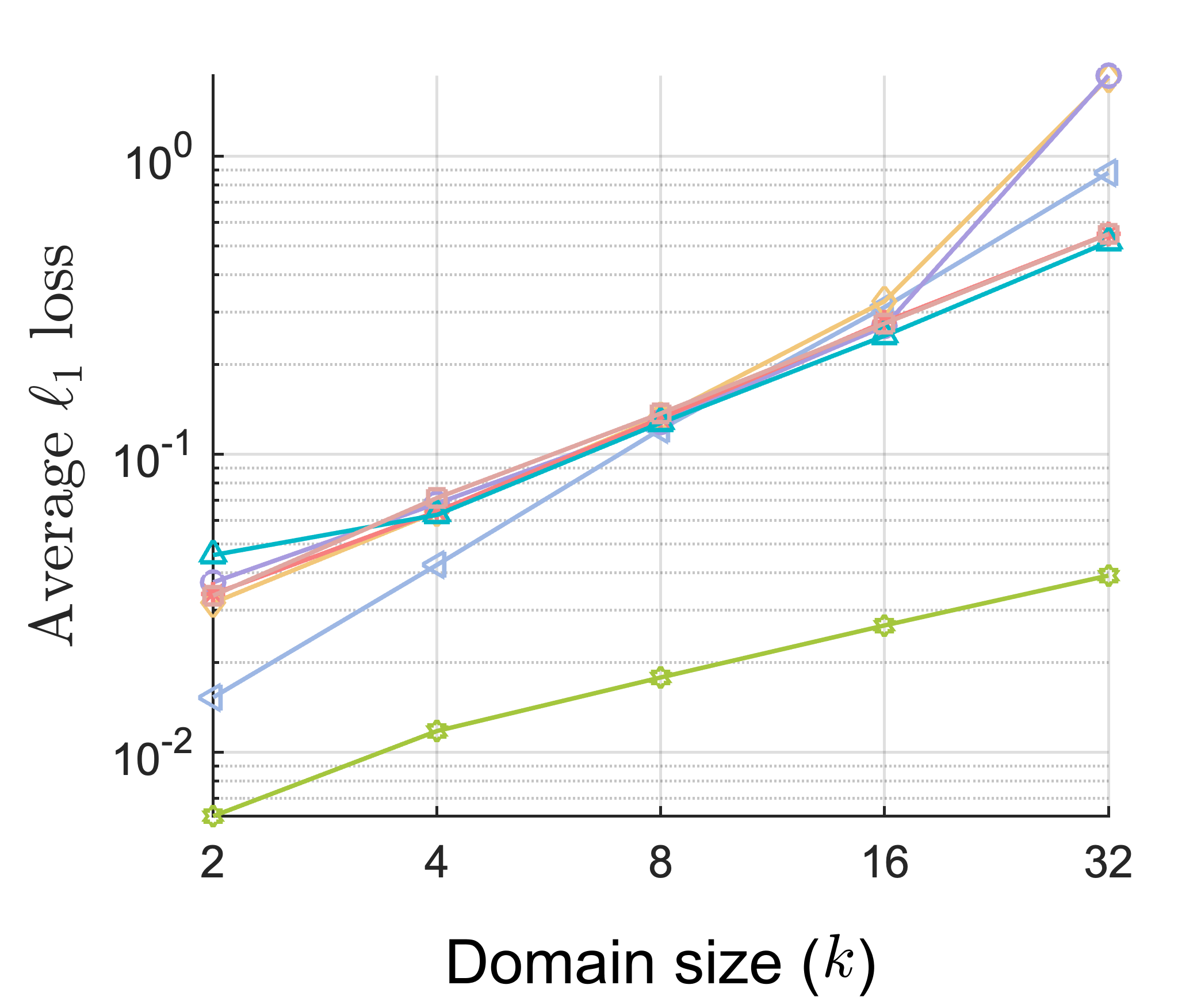}} &
  \subcaptionbox{$\epsilon=2$,\ $M=1000$,\ $t_s=1\,\mathrm{s}$}{\includegraphics[width=\PANELW]{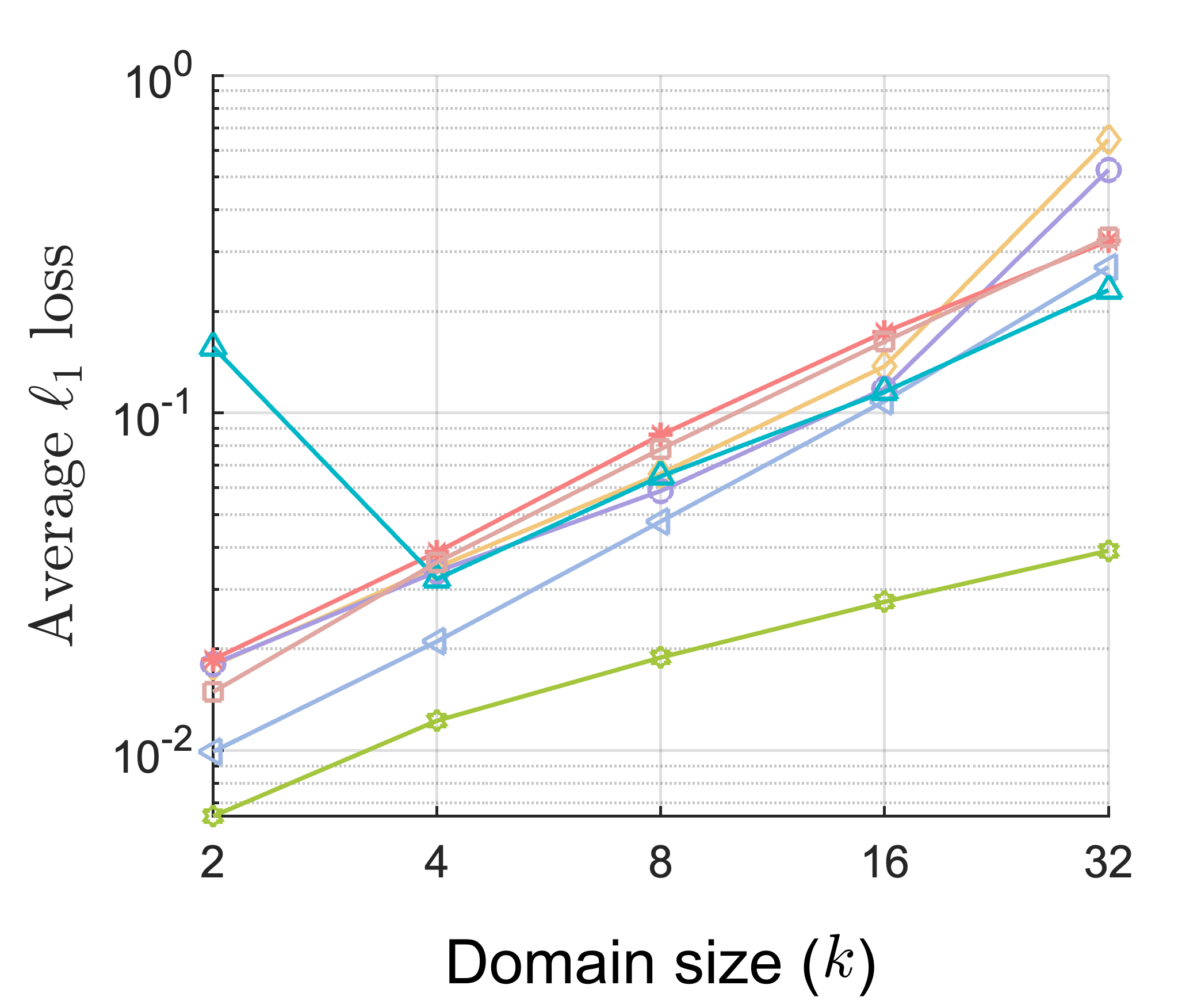}} \\
\end{tabular}
\caption{Average $\ell_1$ losses across methods. Unless stated otherwise, $D=79.4~\mu\mathrm{m}^2/\mathrm{s}$, $r_R=5~\mu\mathrm{m}$, $r_0=10~\mu\mathrm{m}$, $N=10^4$, $\sigma^2=0$.}
\label{fig:all_16_shared_legend_3}
\end{figure*}

At the server, for each candidate $v\in[k]$, let $c_v=\frac{1}{N}\sum_{i=1}^{N}\mathbf{1}\{Y_i\in S_v\}$. Then $\Pr(Y\in S_v\mid X=v)=p$ and for all $x \neq v$ $\Pr(Y\in S_v\mid X = x)=\frac{d}{4}  \frac{2p}{d}+ \frac{d}{4}   \frac{2(1-p)}{d}= \frac{1}{2}$, hence $\mathbb{E}[c_v]=p_v p+(1-p_v)\,\frac12$, and an unbiased estimator is $\hat p_v=\frac{c_v-1/2}{p-1/2}=\frac{2(e^\epsilon+1)}{e^\epsilon-1}\big(c_v-\frac12\big)$.

A primary benefit of HR is its low server-side time complexity, which is achieved through the fast multiplication property of Hadamard matrices. However, this advantage is not strictly necessary in our system model. We assume the central data collector is computationally unconstrained, shifting the focus entirely to the resource limitations of the molecular transmitters.

Regarding implementation, the only LDP component that runs on the constrained node $T_i$ is the local randomizer, which needs just a few random draws per report; the estimators run at the server. The randomness may come from the intrinsic stochasticity of biochemical processes, e.g., gene expression noise in engineered cells \cite{elowitz_stochastic}. This is a simulation-level study; the experimental realization of such a randomizer is left for future work.

\section{Simulation}\label{sec:sim}

\subsection{Simulation Mechanism}\label{sec:sim_mech}

We now compare the LDP mechanisms of Section~\ref{sec:ldp} under the MC channel of Section~\ref{sec:channel} across a broad range of scenarios. Unless stated otherwise, we use $D=79.4~\mu\mathrm{m}^2/\mathrm{s}$, $r_R=5~\mu\mathrm{m}$, $r_0=10~\mu\mathrm{m}$ \footnote{These simulation parameters are extensively used in the MC literature, representing a channel in which each transmitted and detected molecule is a human insulin hormone \cite{ISI_mitigating_methods_2015}.}, a reference channel memory $I_0=200$, and $N=10^4$ users. After fixing $(k,\epsilon)$ and the MC parameters, we generate $100$ random ground-truth distributions $\{\mathbf{p}^{(i)}\}_{i=1}^{100}$ by sampling uniformly from the probability simplex (via a symmetric Dirichlet sampler).

For each user $U_j$ and each distribution index $i$, we draw $X_j^{(i)}\sim \mathbf{p}^{(i)}$ and apply the chosen $\epsilon$-LDP mechanism to obtain a privatized report $Y_{j}^{(i)}\in\mathcal{Y}_m$ (the output alphabet under method $m$), which is then mapped to a binary string $\mathbf{b}^{(i)}_{j,m}:= \phi_m(Y_{j}^{(i)})$ via a method-specific encoding $\phi_m:\mathcal{Y}_m \to \{0,1\}^{\ell_m}$, where $\ell_m$ is the (fixed) report length in bits under method $m$. With $d=\lceil\log_2 k\rceil$, $\ell_m=d$ for KRR, $\ell_m=k$ for Basic RAPPOR and OUE, $\ell_m=d+1$ for BLH, $\ell_m=\lceil\log_2 g^{m+1}\rceil$ for OLH and $\ell_m=\lceil\log_2(k+1)\rceil$ for HR. The unprivatized baseline, denoted by $m=0$, transmits the $d$-bit representation of $x-1$ itself, so that $\ell_0=d$. For BLH and OLH, the binary report contains both the randomized output and the randomly chosen hash vector (Sections \ref{sec:blh} and \ref{sec:olh}).

For each user, we append the $100$ binary reports into a single long binary sequence $\mathbf{b}_{j,m}:=\mathrm{concat}\big(\mathbf{b}^{(1)}_{j,m},\ldots,\mathbf{b}^{(100)}_{j,m}\big)\in \{0,1\}^{100 \cdot \ell_m}$ and transmit it contiguously. This is important because ISI is the major impairment in diffusion-based MC \cite{ISI_mitigating_methods_2015}: molecule arrivals from previous 1-bit transmissions affect the molecule counts in subsequent signal intervals. By transmitting long sequences, we fully capture ISI rather than resetting the channel between reports.

Different LDP mechanisms produce different report lengths and different numbers of 1-bits, both of which directly influence ISI. To ensure a fair comparison, we normalize the signal interval duration, channel-memory length, and the number of molecules emitted per 1-bit so that every method uses the same total transmission time and a comparable (i.e. close) total number of emitted molecules as the unprivatized baseline \cite{moac, normalization}. Let $W_m := \sum_{j=1}^{N}\sum_{t=1}^{100 \cdot \ell_m} b_{j,m}[t]$ be the total number of $1$-bits of all the users accumulated under method $m$, where $b_{j,m}[t]\in\{0,1\}$ denotes the $t$-th bit of $\mathbf{b}_{j,m}$. With $m=0$ denoting the unprivatized baseline, we set \begin{equation}
t_{s,m}
=
t_{s,0}\frac{\ell_0}{\ell_m},
\quad
M_m
=
\left\lfloor
M_0\frac{W_0}{W_m}
\right\rceil,
\quad
I_m
=
\left\lceil
I_0\frac{\ell_m}{\ell_0}
\right\rceil .
\label{eq:normalization}
\end{equation}

This ensures $\ell_m t_{s,m}=\ell_0t_{s,0}$, $M_mW_m\approx M_0W_0$, and $I_m t_{s,m}\approx I_0t_{s,0}$ for all methods. In all figures and captions, $t_s$ and $M$ denote the baseline values $t_{s,0}$ and $M_0$ (i.e., the values of the unprivatized scheme); each method $m$ then operates with the normalized $t_{s,m}$, $M_m$ and $I_m$ (e.g., for $k=16$ and $\epsilon=1$, $t_{s,m}=t_{s,0}$, $\tfrac{4}{5}t_{s,0}$, $\tfrac{4}{7}t_{s,0}$ and $\tfrac{1}{4}t_{s,0}$ for KRR, BLH/HR, OLH and Basic RAPPOR/OUE, respectively). The last normalization gives all methods approximately the same physical memory duration, $I_m t_{s,m}\approx I_0 t_{s,0}=200$~s by default. By Eq.~\eqref{eq:F}, the fraction of emitted molecules that is absorbed only after this window, $F(\infty)-F(I_0t_{s,0})$, is $1.1\%$ for the default parameters, $1.8\%$ at the largest distance $r_0=24~\mu\mathrm{m}$ of Fig.~\ref{fig:all_16_shared_legend_3}(d), $3.2\%$ at $r_0=24~\mu\mathrm{m}$ with $t_{s,0}=0.3$~s (Fig.~\ref{fig:all_16_shared_legend}(d,h)), and at most $3.5\%$ for the shortest window ($t_{s,0}=0.1$~s); these late arrivals form a nearly constant floor absorbed by the detection threshold and, being the same for all methods, do not favor any scheme.

Given the transmitted bit sequence, we simulate the received molecule-count sequence using \eqref{eq:Nh} with the multinomial channel model of Section~\ref{sec:channel}. We then apply threshold detection: a transmitted bit $b_{j,m}[t]$ is detected as $\hat b_{j,m}[t]=1$ iff the observed molecule count in the corresponding interval is larger than or equal to a fixed threshold $\tau \in \mathbb{N}$. To ensure a consistent comparison across privacy schemes, we choose a single global threshold per method by minimizing the total BER aggregated across all users, $\tau_m^\star \in \arg\min_{\tau} \sum_{j=1}^{N} \mathrm{Ham}\big(\hat{\mathbf{b}}_{j,m}(\tau),\mathbf{b}_{j,m}\big)$. After detection, we split each user’s detected bitstream back into $100$ reports, invert the method-specific binary mapping\footnote{Because the MC channel can flip bits, the detected binary word may not correspond to any valid report (e.g., for OLH it may decode to $\hat z\notin\{0,\ldots,g^{m+1}-1\}$). In this case, we replace it by a uniformly random valid report before running the corresponding estimator.}, and apply the corresponding unbiased estimator from Section~\ref{sec:ldp} to obtain $\hat{\mathbf{p}}^{(i)}_m$ for each $i$. Finally, we compute the average $\ell_1$ error $\ell_1^m := \frac{1}{100}\sum_{i=1}^{100}\|\mathbf{p}^{(i)}-\hat{\mathbf{p}}^{(i)}_m\|_1$.

The estimators of Section~\ref{sec:ldp} are unbiased when the server receives the transmitted reports $Y$; in our system it receives the detected reports $\hat Y$ instead. Averaged over the transmitted stream, the MC link acts approximately as a memoryless transition matrix $C$ with entries $C_{y,\hat y}=\Pr(\hat Y=\hat y\,|\,Y=y)$ (ISI makes the detected report also depend on the neighbouring reports, which this average absorbs), so the server effectively observes the mechanism $QC$. Privacy is unaffected: $QC$ is a post-processing of $Q$ and is therefore still $\epsilon$-LDP, whatever $C$ is.

The estimates, however, become biased. All estimators of Section~\ref{sec:ldp} have the form $\hat p_v=(c_v-q)/(p-q)$, where $c_v$ is the observed frequency of a mechanism-specific event (e.g., $\{Y=v\}$ for KRR) that occurs with probability $p$ when $X=v$ and $q$ otherwise, with $q=\tfrac12$ for BLH and HR and $q=1/g$ for OLH; hence $\mathbb{E}[c_v]=p_v p+(1-p_v)q$ and $\mathbb{E}[\hat p_v]=p_v$. Channel errors change the two probabilities to some $(\tilde p,\tilde q)$; for simplicity of exposition we take them to be the same for all $x$, which only serves to display the form of the deviation and does not affect the simulations, which use the full channel.

Then $\mathbb{E}[c_v]=p_v\tilde p+(1-p_v)\tilde q$, while the server still divides by the original constants. Substituting gives \begin{equation}\label{eq:bias}
\mathbb{E}[\hat p_v]-p_v=\frac{(\tilde p-\tilde q)-(p-q)}{p-q}\,p_v+\frac{\tilde q-q}{p-q},
\end{equation} i.e., $\mathbb{E}[\hat p_v]=\frac{\tilde p-\tilde q}{p-q}\,p_v+\frac{\tilde q-q}{p-q}$. A channel-aware estimator that calibrates $(\tilde p,\tilde q)$ from pilot transmissions would remove this bias at the cost of a larger variance, and is left for future work; here the server does not know the channel and we report the uncorrected estimators; the $\ell_1$ errors in Figs.~\ref{fig:all_16_shared_legend_3} and \ref{fig:all_16_shared_legend} therefore include this bias.

\subsection{Evaluation of Simulation Results}\label{sec:eval}

Fig.~\ref{fig:all_16_shared_legend_3} reports the average $\ell_1$ error across the considered scenarios. Fig.~\ref{fig:all_16_shared_legend_3}(a) shows the performance across different privacy budgets ($\epsilon$). All methods improve with $\epsilon$, except that OLH (which is equivalent to BLH for $\epsilon=0.125$ and $0.25$) loses its advantage for $\epsilon\ge2$ and its error even increases at $\epsilon=8$, since its report length grows with $g\approx e^{\epsilon}+1$ and its normalized signal interval shrinks accordingly, and BLH levels off for $\epsilon\ge4$ because it hashes to a single bit. We focus on two representative privacy budgets, $\epsilon \in \{1,2 \}$, which are commonly used as benchmark values in empirical evaluations of LDP mechanisms \cite{2-oue-4-olh-7-blh}.

\begin{figure*}[!t]
\centering
\setlength{\PANELW}{0.243\textwidth}
\setlength{\ROWGAP}{0.5mm}
\begin{tabular}{@{}c@{}c@{}c@{}c@{}}
\multicolumn{4}{@{}c@{}}{
  \includegraphics[width=4\PANELW]{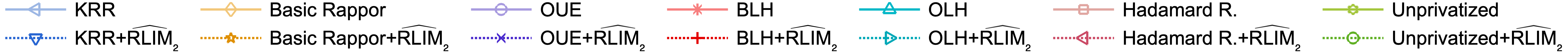}
}\\[-1mm]
  \subcaptionbox{$\epsilon=1$, $M=100$}{\includegraphics[width=\PANELW]{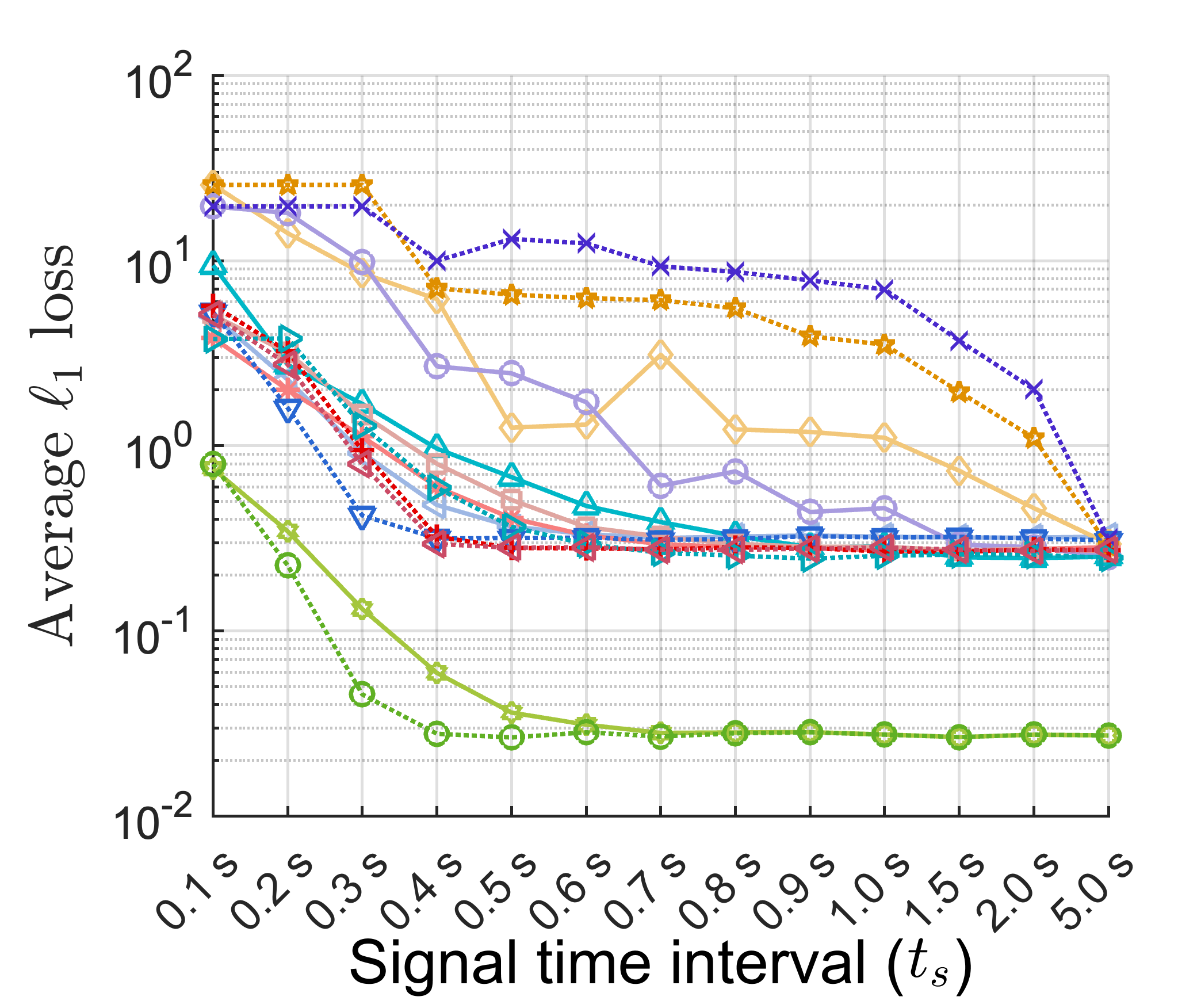}} &
  \subcaptionbox{$\epsilon=1$, $t_s=0.3$ $\mathrm{s}$}{\includegraphics[width=\PANELW]{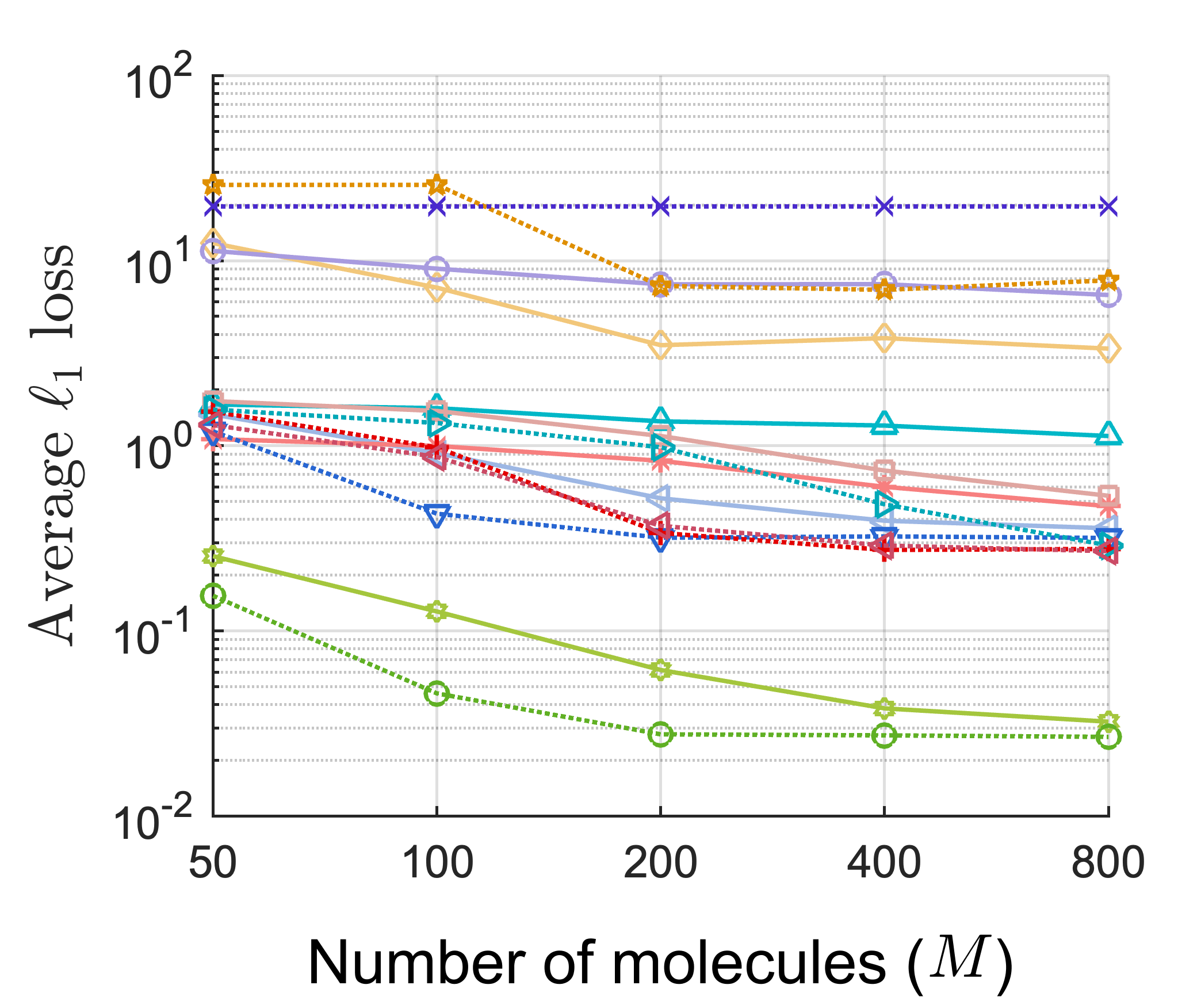}} &
  \subcaptionbox{$\epsilon=1$, $M=100, t_s=1$ $\mathrm{s}$}{\includegraphics[width=\PANELW]{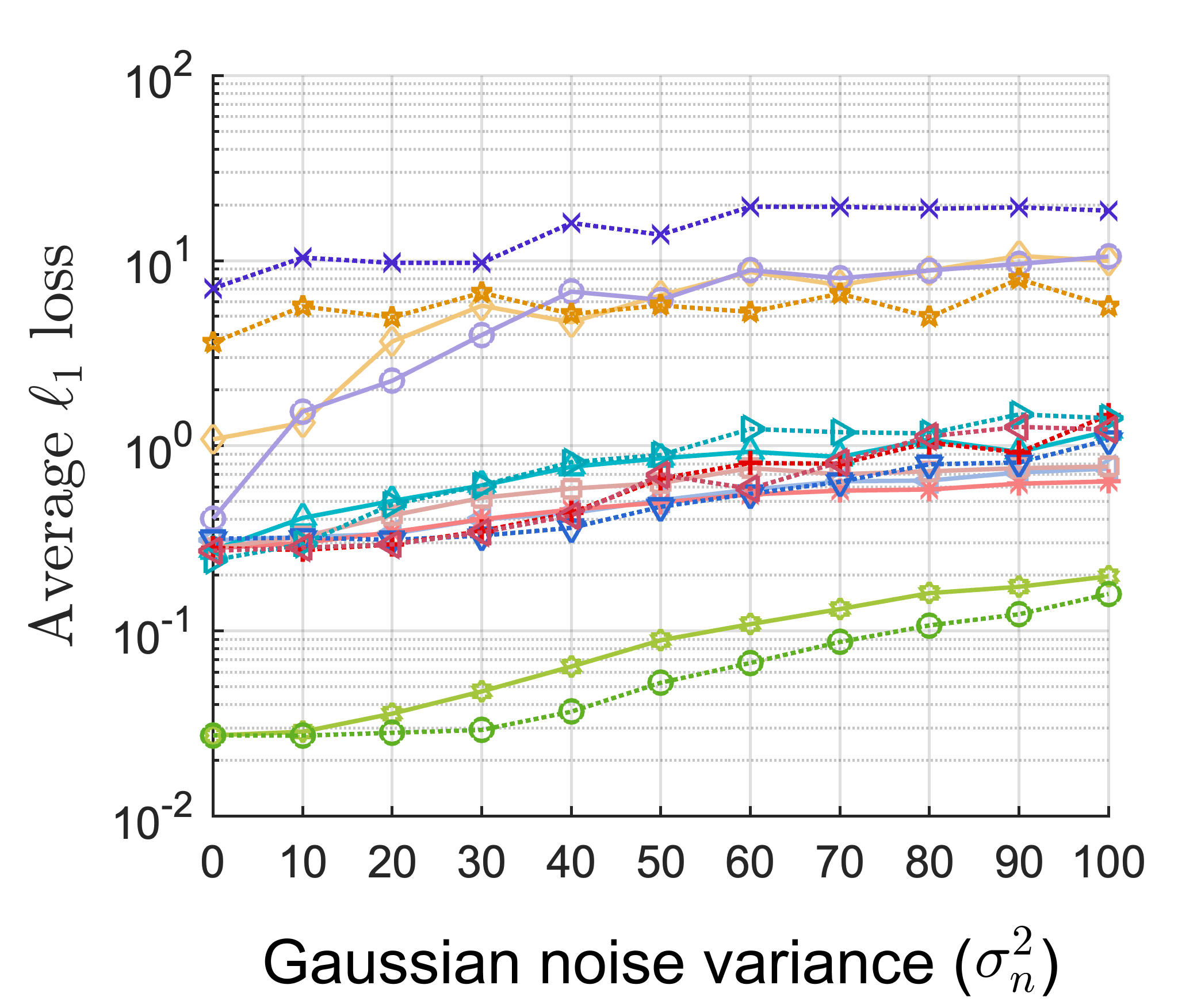}} &
  \subcaptionbox{$\epsilon=1$, $t_s=0.3$ $\mathrm{s}$, $M=100$ }{\includegraphics[width=\PANELW]{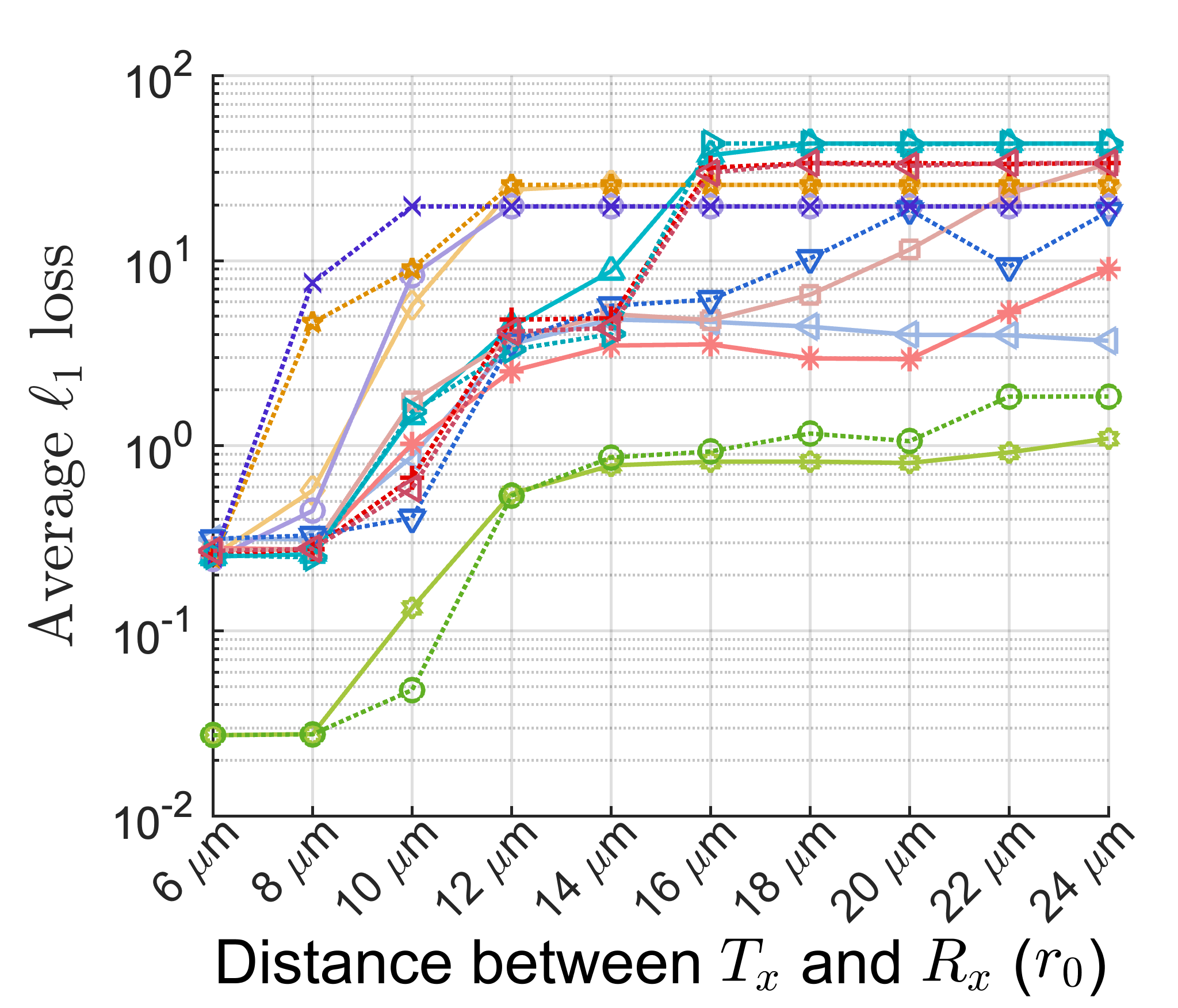}}
  \\[\ROWGAP]
  \subcaptionbox{$\epsilon=2$, $M=100$}{\includegraphics[width=\PANELW]{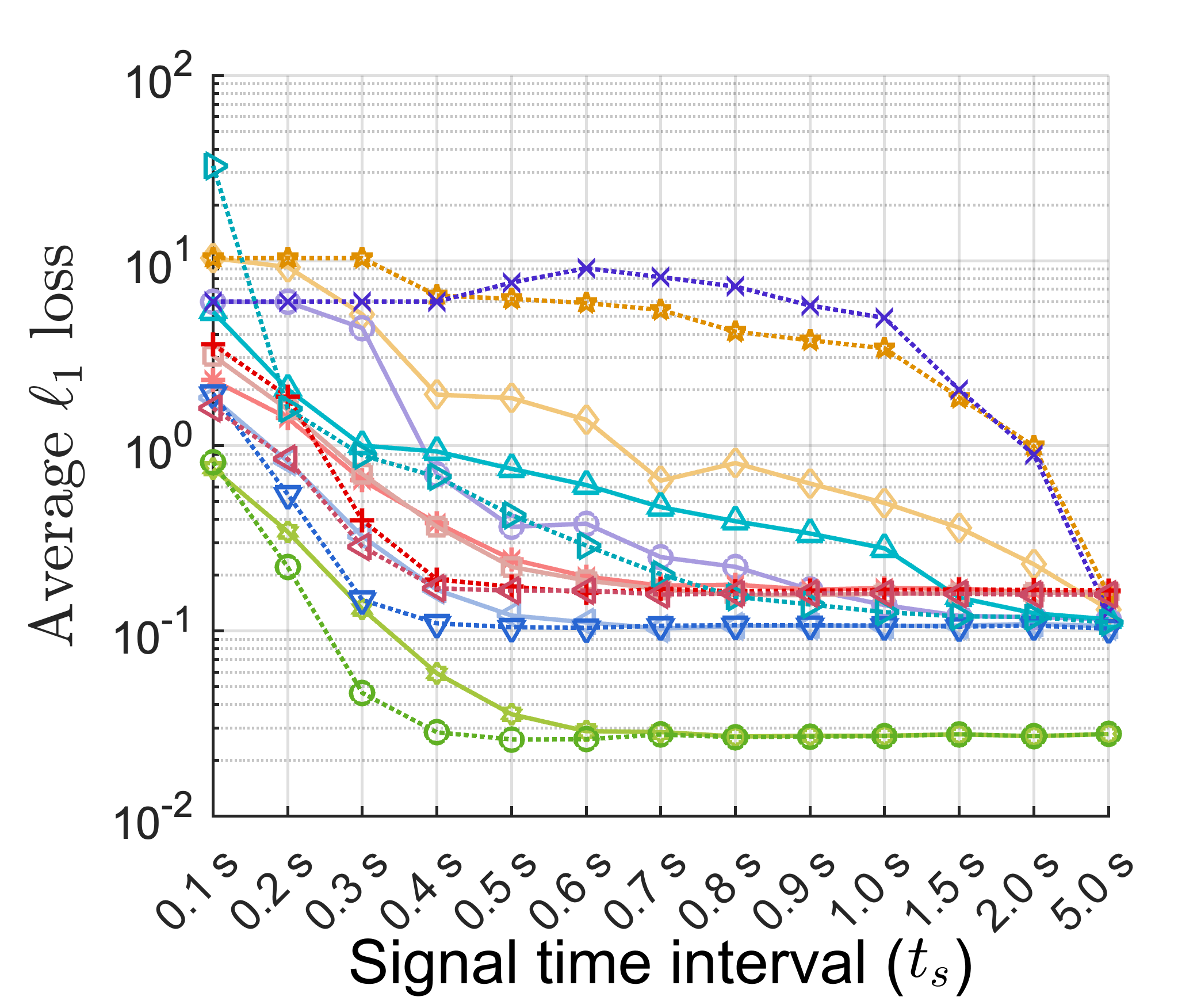}} &
  \subcaptionbox{$\epsilon=2$, $t_s=0.3$ $\mathrm{s}$}{\includegraphics[width=\PANELW]{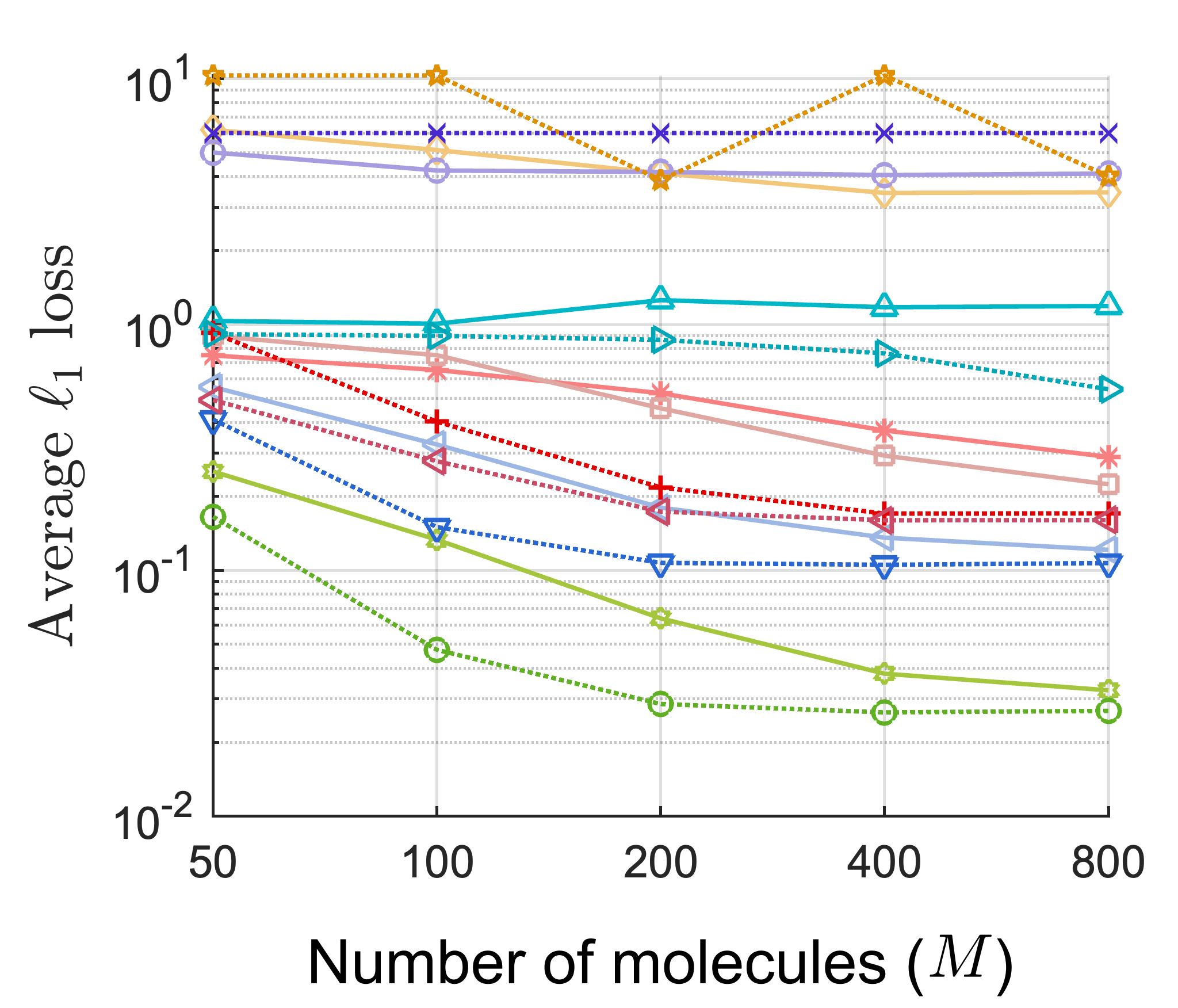}} &
  \subcaptionbox{$\epsilon=2$, $M=100, t_s=1$ $\mathrm{s}$}{\includegraphics[width=\PANELW]{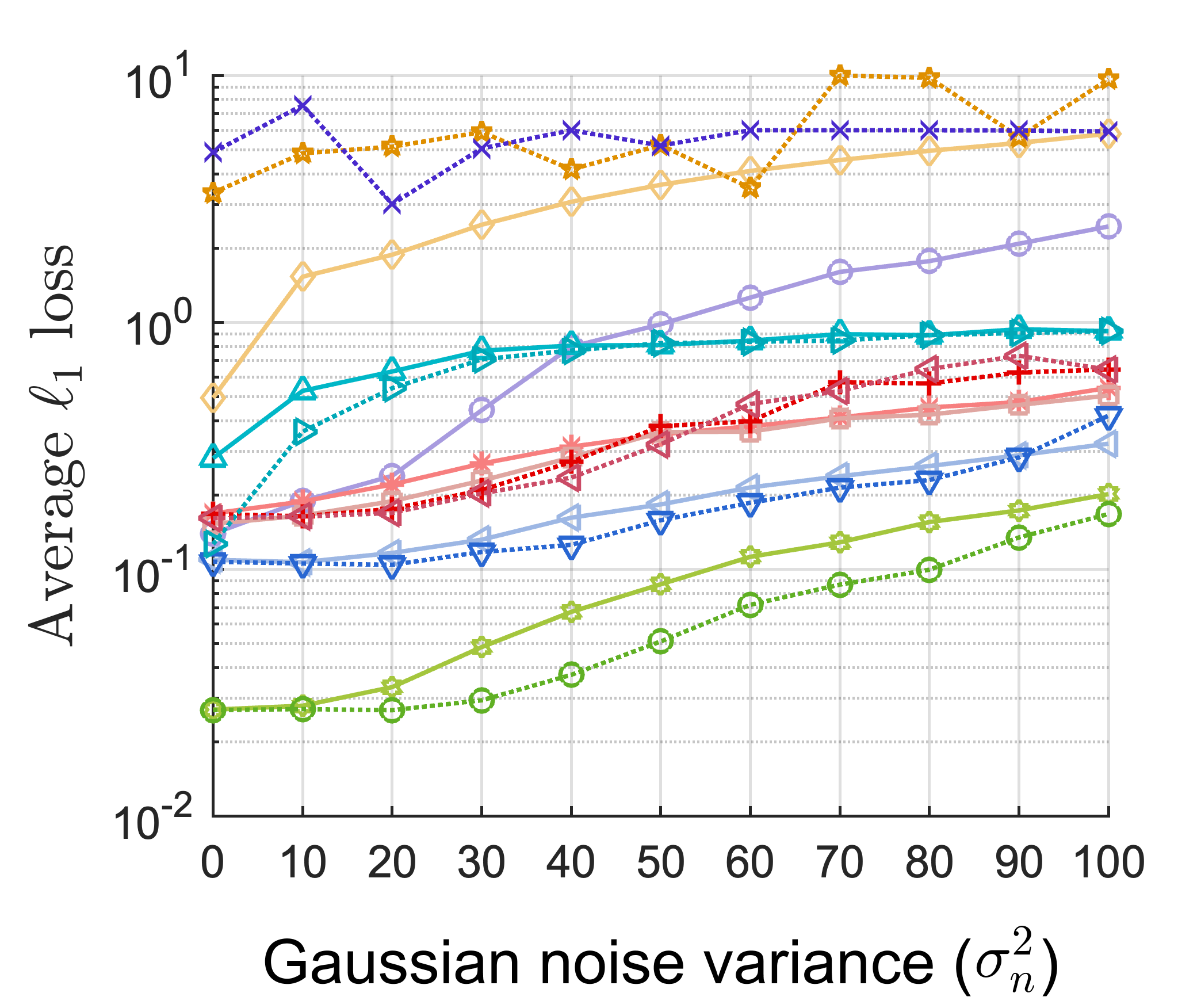}} &
  \subcaptionbox{$\epsilon=2$, $t_s=0.3$ $\mathrm{s}$, $M=100$  }{\includegraphics[width=\PANELW]{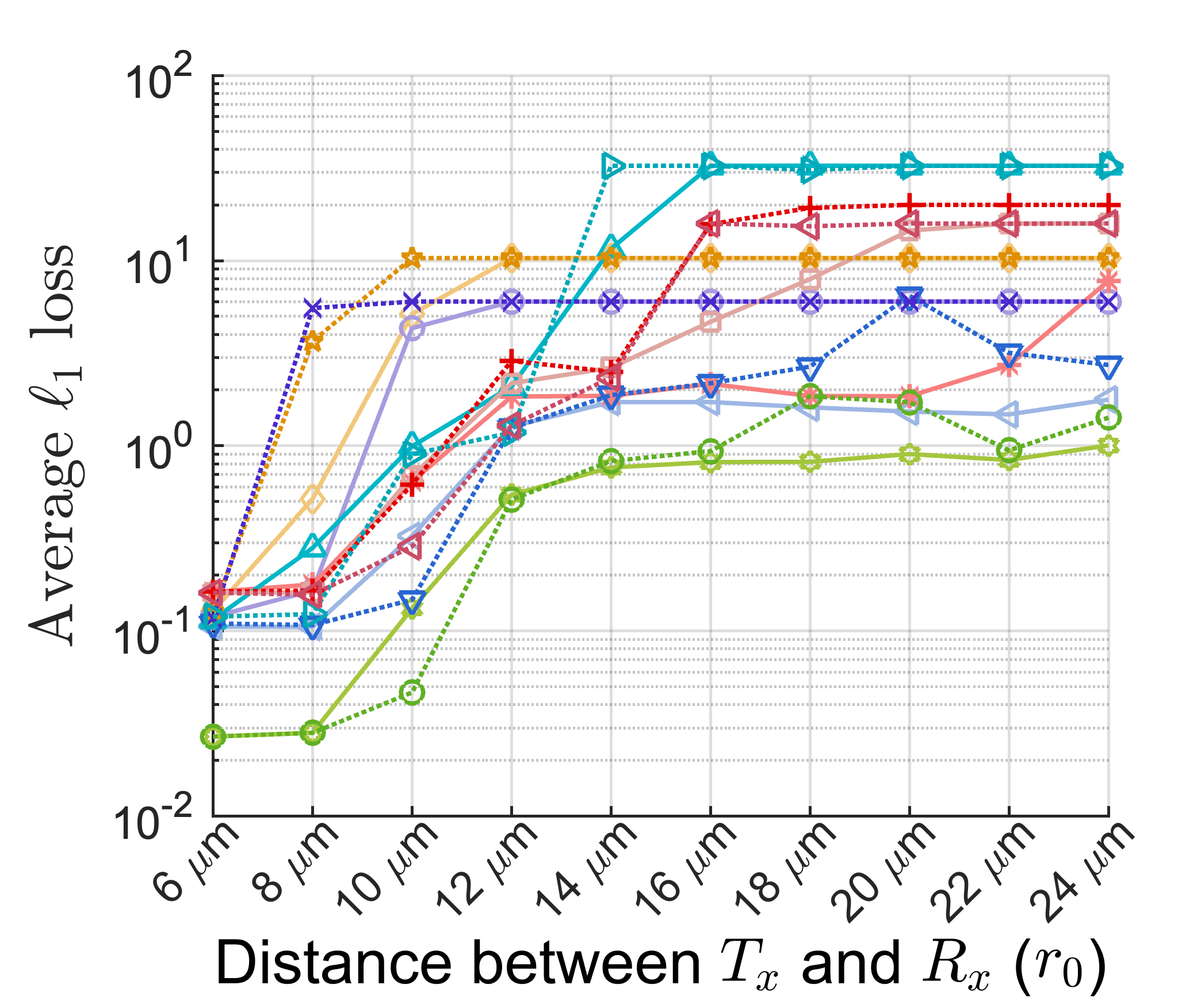}} \\
\end{tabular}
\caption{Average $\ell_1$ losses across methods. Unless stated otherwise, $k=16$, $D=79.4~\mu\mathrm{m}^2/\mathrm{s}$, $r_R=5~\mu\mathrm{m}$, $r_0=10~\mu\mathrm{m}$, $N=10^4$, $\sigma^2=0$.}
\label{fig:all_16_shared_legend}
\end{figure*}

For $\epsilon=1$, Fig.~\ref{fig:all_16_shared_legend_3}(g) shows that KRR is the best method for small domains ($k\le8$), whereas for $k \ge 16$, when $t_s$ and $M$ are sufficiently large, OLH becomes the most advantageous method, with BLH and HR close behind and KRR falling behind them. However, as shown in Fig.~\ref{fig:all_16_shared_legend_3}(b,c,d), when the channel quality deteriorates due to a smaller $t_s$, lower $M$, or a larger transmitter-receiver distance, KRR and BLH become the most advantageous methods while OLH falls behind: KRR has the lowest error for $t_s\in\{0.2,0.3\}$~s in Fig.~\ref{fig:all_16_shared_legend_3}(b) and for $14\le r_0\le18~\mu\mathrm{m}$ in Fig.~\ref{fig:all_16_shared_legend_3}(d), whereas BLH has the lowest error at $M=50$ in Fig.~\ref{fig:all_16_shared_legend_3}(c) as well as at $t_s=0.1$~s and for $r_0\ge20~\mu\mathrm{m}$, where, however, the overall reliability of all methods remains poor ($\ell_1>5$). This robustness is consistent with their short reports: the report lengths ($\ell_{\mathrm{KRR}}=\ell_0<\ell_{\mathrm{BLH}}=\ell_{\mathrm{HR}}<\ell_{\mathrm{OLH}}<\ell_{\mathrm{RAPPOR}}=\ell_{\mathrm{OUE}}$) determine the normalized $t_{s,m}$ directly, and the numbers of transmitted 1-bits $W_m$, which happen to follow the same ordering, determine the normalized $M_m$ under Eq.~\eqref{eq:normalization}. For $\epsilon=2$, Fig.~\ref{fig:all_16_shared_legend_3}(h) shows the same trend: OLH outperforms KRR at $k=32$, whereas KRR is better for $k\le16$; the $\epsilon=2$ sweeps over $t_s$, $M$ and $r_0$ at $k=32$ (not shown) again favor KRR whenever the channel degrades. Moreover, for $\epsilon=2$ KRR outperforms BLH under all simulated channel conditions (see also the uncoded curves in Fig.~\ref{fig:all_16_shared_legend}(e--h)), since the single-bit BLH gains little from the larger privacy budget (Fig.~\ref{fig:all_16_shared_legend_3}(a)).

Regarding the population size parameter $N$, Fig.~\ref{fig:all_16_shared_legend_3}(e) shows that the estimation error of every method decreases as $N$ increases and that OLH remains the most accurate throughout; the unary schemes, Basic RAPPOR in particular, improve more slowly because their $16$-bit reports suffer the most channel errors, so that the channel-induced bias of Section~\ref{sec:sim_mech}, which does not vanish with $N$, dominates. We evaluate the impact of Gaussian counting noise variance ($\sigma^2$) in Fig.~\ref{fig:all_16_shared_legend_3}(f) for $M=100$: BLH degrades most gracefully, followed by KRR and HR, OLH degrades more strongly, and the unary schemes, which receive the smallest normalized molecule budget, degrade sharply. A larger molecule budget per 1-bit ($M=1000$, not shown) significantly mitigates the adverse effects of counting noise.

In summary, at distances of $r_0 \le 12~\mu\mathrm{m}$, OLH typically yields the lowest estimation error when the unprivatized signal interval is approximately $t_s \ge 0.5\,\mathrm{s}$ (given $M \ge 1000$, Fig.~\ref{fig:all_16_shared_legend_3}(b)) or $t_s \ge 1.5\,\mathrm{s}$ (given $M \ge 100$, uncoded curves in Fig.~\ref{fig:all_16_shared_legend}(a)); when the available channel resources fall below these ranges, KRR and BLH generally provide more robust alternatives.

\section{Run Length Limited ISI Mitigation LDP Coding (RLIM-LDP)} \label{sec:mc_channel_coding_ldp}

To enhance the reliability of the private MC channel, we propose \textit{RLIM-LDP}, a unified framework integrating LDP with the RLIM coding family introduced in \cite{MC_channel_coding}. While LDP mechanisms ensure privacy by randomizing user data, they do not inherently mitigate physical channel impairments such as ISI \cite{ISI_mitigating_methods_2015}. By encoding privatized reports via RLIM, we minimize the frequency of molecular emissions, thereby optimizing the energy budget and reducing the final distribution estimation error $\ell_1$ at the central server.

For every mechanism considered, the output can be represented as a vector of fixed length $\ell_m$ over a finite integer alphabet $\mathcal{A}_B = \{0, 1, \dots, B-1\}$, where $B=g$ for OLH and $B=2$ for all other mechanisms. We define the bijective indexing map $\phi_m: \mathcal{A}_B^{\ell_m} \to \{0, 1, \dots, B^{\ell_m} - 1\}$ as the positional interpretation $\phi_m(\mathbf{y}) = \sum_{j=1}^{\ell_m} y_j \cdot B^{\ell_m - j}$ of the output vector $\mathbf{y} = (y_1, \dots, y_{\ell_m}) \in \mathcal{Y}_m$, which assigns a unique integer index $s$ to each privatized report, and let $S_m = B^{\ell_m}$ denote the cardinality of the report space.

To transmit these $S_m$ privatized reports more reliably over the MC channel, we map them into a codebook $\mathcal{C}_m \subset \{0,1\}^{n_m}$ using the enhanced version of RLIM coding outlined in Section IV.B.5 in \cite{MC_channel_coding}, rather than the standard binary representations that are vulnerable to ISI. The core principle of RLIM is to enforce a minimum temporal spacing between molecular emissions while minimizing the overall frequency of those emissions. Let $\widehat{RLIM}_{2}(n_m)$ denote the space of all length-$n_m$ binary sequences satisfying a $(2, \infty)$-run-length constraint, meaning every 1-bit is followed by at least two 0-bits; the hat notation \cite{MC_channel_coding} indicates that the all-zero vector is permitted.

For each LDP method $m$, we construct a corresponding channel codebook, defined as $\mathcal{C}_m \subset \widehat{RLIM}_{2}(n_m)$, by first determining the minimum codeword length $n_m$ required to satisfy the cardinality condition $|\widehat{RLIM}_{2}(n_m)| \ge S_m$ (for $k=16$ and $\epsilon=1$: $n_m=9$ for KRR, $11$ for BLH and HR, $13$ for OLH and $31$ for Basic RAPPOR and OUE, whose report space has $2^{k}$ elements). From this feasible space, $\mathcal{C}_m$ is formed by selecting the $S_m$ codewords with the lowest Hamming weight: as established in \cite{MC_channel_coding}, a sparser codebook minimizes the total number of transmitted 1-bits, which under the molecule-budget normalization allows a larger molecule budget per 1-bit and thus reduces BER. The transmitter maps the privatized report $\mathbf{y}$ to its index $s = \phi_m(\mathbf{y})$ and transmits the corresponding codeword $\mathbf{c}_s \in \mathcal{C}_m$.

\captionsetup[subfigure]{labelformat=parens} \captionsetup[subfigure]{justification=centering, singlelinecheck=false}

\subsection{RLIM-LDP Simulation Results}\label{sec:rlim_results}

The data generation and physical channel configurations are identical to Section~\ref{sec:sim}. To ensure a fair comparison with the unprivatized baseline, we apply the same time, molecule-budget, and physical channel-memory normalizations as in Section~\ref{sec:sim}. For an RLIM-coded method, the actually transmitted report has codeword length $n_m$. Therefore, the signal interval and channel memory are normalized according to \begin{equation}
t_{s,m}^{\mathrm{RLIM}}
=
t_{s,0}\frac{\ell_0}{n_m},
\qquad
I_m^{\mathrm{RLIM}}
=
\left\lceil
I_0\frac{n_m}{\ell_0}
\right\rceil .
\end{equation} The molecule budget is normalized as in Eq.~\eqref{eq:normalization}, with $W_m$ computed from the actually transmitted RLIM codewords. Hence, the total transmission time and molecule budget remain comparable to the unprivatized baseline, while $I_m^{\mathrm{RLIM}}t_{s,m}^{\mathrm{RLIM}}
\approx I_0t_{s,0}$ preserves approximately the same physical channel-memory duration.

At the receiver, absorbed molecules are counted and mapped to a binary sequence using a fixed threshold detector. Because threshold optimization over the full population is computationally expensive due to the end-to-end decoding pipeline of RLIM codes, we instead calibrate the threshold on a pilot set $\mathcal{K}$ comprising the first $100$ users’ transmissions, minimizing the end-to-end symbol error rate (SER): $\tau^\star = \operatorname*{argmin}_{\tau} \sum_{k \in \mathcal{K}} \mathbf{1}\{ \hat{s}_k(\tau) \neq s_k \}$, where $\hat{s}_k(\tau)$ denotes the decoded index of the $k$-th pilot transmission under threshold $\tau$. The detected sequence is then processed by the linear-time RLIM error correction algorithm, which enforces the $(2, \infty)$ run-length constraint and is equivalent to a Viterbi decoder with the deterministic last-wins rule \cite{MC_channel_coding}. The corrected sequence is decoded to the estimated index $\hat{s}$, which is mapped back to the LDP report $\hat{\mathbf{y}} = \phi_m^{-1}(\hat{s})$ and passed to the central server for aggregation.

For the privacy budgets $\epsilon \in \{1,2\}$, the RLIM-LDP simulation results are presented in Fig.~\ref{fig:all_16_shared_legend}. We first observe in Fig.~\ref{fig:all_16_shared_legend}(a, e) that, with the exception of Basic RAPPOR and OUE, RLIM coding typically improves the average $\ell_1$ loss when the signal interval duration is small: the gains are most pronounced for $t_s \in \{0.2,0.3,0.4\}\,\mathrm{s}$ and remain visible up to $t_s=0.6\,\mathrm{s}$ (up to $1$~s for OLH, and $1.5$~s at $\epsilon=2$); for $t_s\le0.2$~s, however, coded BLH is even worse than uncoded BLH. Basic RAPPOR and OUE do not benefit because their codeword length $n_m=31$ makes their signal interval $31/4$ times shorter than $t_{s,0}$. Fig.~\ref{fig:all_16_shared_legend}(b, f) shows that at $t_s=0.3$~s the coded KRR, OLH and HR outperform their uncoded counterparts for every $M$, and coded BLH for $M\ge200$ ($M\ge100$ at $\epsilon=2$).

At $t_s=1$~s (Fig.~\ref{fig:all_16_shared_legend}(c, g)), the RLIM-coded KRR, BLH, OLH and HR mostly retain a modest advantage over their uncoded counterparts under moderate counting noise (up to $\sigma^2\approx30$ for $\epsilon=1$ and $\sigma^2\approx40$ for $\epsilon=2$), but this advantage vanishes at larger $\sigma^2$, where the coded schemes are mostly worse than the uncoded ones. Fig.~\ref{fig:all_16_shared_legend}(d, h) shows that RLIM coding improves reliability at $r_0=10~\mu\mathrm{m}$ (most clearly for KRR and HR; OLH benefits only at $\epsilon=2$); at $r_0=12~\mu\mathrm{m}$, gains remain visible for OLH (and for HR at $\epsilon=2$), whereas for $r_0\ge14~\mu\mathrm{m}$ the coded schemes are generally worse than the uncoded ones, although in that regime neither is useful ($\ell_1>1$).

Overall, RLIM channel coding provides a noticeable performance advantage to the most competitive LDP mechanisms under limited channel resources. Within our simulated parameters, this improvement is particularly significant for short signal intervals ($0.2\,\mathrm{s}\le t_s \le 0.5\,\mathrm{s}$), moderate communication distances ($r_0 \le 10~\mu\mathrm{m}$), and moderate receiver counting-noise variance.

\section{Conclusion} \label{sec:conclusion}

This paper investigated the integration of LDP into diffusion-based MC networks. By benchmarking state-of-the-art LDP mechanisms under realistic channel conditions, we demonstrated that mechanisms with higher statistical accuracy often generate longer binary reports, which exacerbate ISI and degrade communication reliability. Consequently, our findings indicate that while OLH minimizes distribution-estimation error when channel resources are ample and the alphabet size is moderate to large, the shorter report lengths of the KRR mechanism (and, for $\epsilon=1$, BLH) provide superior robustness when transmission conditions degrade.

To maintain these privacy guarantees under severe nanoscale physical constraints, we proposed the RLIM-LDP architecture. Extensive simulations demonstrated that encoding privatized LDP reports with RLIM coding effectively yields significant improvements in end-to-end reliability and estimation accuracy under tight time and molecule resource constraints. Overall, this work demonstrates that private, population-scale aggregate data analysis can be reliably achieved in diffusive biological environments under practical in-body MC channel conditions. The experimental realization of the local randomizer in a biological transmitter is left for future work.

\bibliographystyle{IEEEtran}

\bibliography{References}
\end{document}